\pdfoutput=1
\documentclass[final,times,1p]{elsarticle}

\usepackage{graphics}
\usepackage{graphicx,stfloats}
\usepackage{color}

\usepackage{amssymb}
\usepackage{amsmath}
\usepackage{amsthm}
\usepackage{amsmath}
\usepackage{soul}
\usepackage{xcolor}
\usepackage{commath}
\usepackage{latexsym}
\usepackage{subfig}
\usepackage{float}
\usepackage{url}
\usepackage{setspace}
\usepackage{multirow}
\usepackage{threeparttable}
\usepackage{booktabs}
\usepackage{textgreek}

\newcommand{\pde}[2]{\frac{\partial {#1}}{\partial {#2}}}   

\newcommand{\ode}[2]{\dfrac{\text{d} {#1}}{\text{d} {#2}}}  

\newcommand{\intl}[2]{\int\limits_{#1}^{#2}}				

\biboptions{sort&compress}

\journal{Combustion and Flame}

\begin{document}
	
\begin{frontmatter}
	
\title{Dynamics of Hydrogen-Oxygen-Argon Cellular Detonations with a Constant Mean \textcolor{black}{Lateral Strain Rate}}

\author[add1]{Q. Xiao\corref{cor1}}
\ead{qxiao067@uottawa.ca}
\author[add1]{M. I. Radulescu}

\address[add1]{Department of Mechanical Engineering, University of Ottawa, Ottawa, \\Ontario K1N 6N5, Canada}
\cortext[cor1]{Corresponding author:}

\begin{abstract}
	
The present work revisits the problem of modelling the real gaseous detonation dynamics at the macro-scale by simple steady one-dimensional (1D) models. Experiments of detonations propagating in channels with exponentially expanding cross-sections were conducted in the H${_2}$/O${_2}$/Ar reactive system.  Steady detonation waves were obtained at the macro-scale, with cellular structures characterized by reactive transverse waves.  For all the mixtures studied, the dependence of the mean detonation speed was found to be in excellent agreement with first principles predictions of quasi-1D detonation dynamics with lateral \textcolor{black}{strain} rate predicted from detailed chemical kinetic models.  This excellent agreement departs from the earlier experiments of Radulescu and Borzou (2018) in more unstable detonations.  The excellent agreement is likely due to the much longer reaction zone lengths of argon diluted hydrogen-oxygen detonations at low pressures, as compared with the characteristic induction zone lengths.  While the cellular instability modifies the detonation induction zone, the detonation dynamics at the macro-scale are arguably controlled by its hydrodynamic thickness. Near the limit, minor discrepancy is observed, with the experimental detonations typically continuing to propagate to slightly higher lateral strain rates and higher velocity deficits. 

\end{abstract}


\end{frontmatter}

\section{Introduction}
\label{Introduction}
Real detonations in gases have been experimentally observed to travel at a speed smaller than the ideal Chapman-Jouguet (CJ) detonation speed by a velocity deficit, due to the presence of non-ideal effects \cite{fickett1979detonation, lee_2008}. These non-ideal factors include lateral \textcolor{black}{flow divergence}, unsteadiness, and momentum and heat losses \cite{fickett1979detonation}. Extensive efforts have been made to quantitatively compare the experimentally measured velocity deficits and propagation limits with the theoretical predictions made by relatively simple models, which build up on the classical  one-dimensional (1D) Zeldovich-von Neumann-Doering (ZND) model \cite{zeldovich1950theory,fay1959two,dove1974velocity, murray1985influence,agafonov1994computation,ISHII20022789, chao2009detonability,kitano2009spinning,camargo2010propagation, ishii2011detonation, zhang2015detonation, gao2016experimental,ZHANG2017193,radulescu2002failure}. The multi-dimensional transient cellular structures, consisting of an intricate ensemble of interacting triple points, shear layers, and transverse waves, of the real gaseous detonations, however, greatly complicate these attempts. They give rise to substantial deviations from the classical 1D ZND detonation structures \cite{lee_2008,SHEPHERD200983}. A significant question hence comes up that, can the extended ZND model, which neglects the time varying cellular structures, be able to model the real detonation dynamics in the presence of losses at the macro-scale? The present paper addresses this question.

This question has also been attempted in the past by investigating detonation propagation in narrow tubes \cite{ISHII20022789, chao2009detonability,camargo2010propagation, ishii2011detonation,zhang2015detonation, gao2016experimental,ZHANG2017193} and tubes with porous walls  \cite{radulescu2002effect, radulescu2003propagation}. The mean propagation velocities of detonations under varied initial pressures as well as the propagation limits were experimentally determined for various mixtures. Due to the lateral \textcolor{black}{flow divergence} from the growth of the viscous boundary layer on tube walls acting as a mass sink \cite{fay1959two} or from the porosity of the walls,  streamlines in the steady reaction zone are diverged resulting in a globally curved detonation front experiencing a velocity deficit \cite{chinnayya2013computational}. In narrow tubes, the boundary layer theory of Fay \cite{fay1959two} was adopted for evaluating the global \textcolor{black}{lateral strain rate}; while in porous tubes, where constant \textcolor{black}{flow divergence} was assumed, it was estimated from the permeability of the porous wall by assuming a choked flow \cite{radulescu2003propagation}. The generalized ZND model with lateral \textcolor{black}{strain} was then applied to model the detonation dynamics. The authors have found that, the experimentally obtained detonation velocity deficits and propagation limits are in generally good agreement with the theoretical predictions, made with the steady ZND model for weakly unstable detonations, which are characterized by regular cellular structures; while for the unstable detonations with irregular cells, the agreement is poor. Nevertheless, a number of simplifying assumptions and matching constants were made in these works for the predictions. Firstly, there exist limitations in the unrealistic assumption of uniform flow divergence for detonations in narrow and porous tubes, as  Chinnayya et al. \cite{chinnayya2013computational} and Mazaheri et al. \cite{mazaheri2015experimental} have numerically demonstrated that a curved detonation front with \textcolor{black}{flow divergence} due to wall boundary layers or permeability is not expected to have a unique curvature. Moreover, Fay-type models \cite{ISHII20022789, ishii2011detonation,zhang2015detonation, gao2016experimental,ZHANG2017193} require the empirical inputs of a specifically defined value of pressure ratio $\epsilon$, and a particular length scale for modelling the flow divergence rate, whose impact on the predictions has not been evaluated. All these factors thus potentially diminish the values of the comparisons in relevant works.

Very recently, Radulescu and Borzou \cite{Matei-Ramp} experimentally formulated a novel solution allowing for making a meaningful comparison of the experimental results with theoretical models. Their experimental technique involved two exponentially shaped channels. The constant logarithmic derivative of the cross-sectional area enabled detonations to propagate with a constant mean \textcolor{black}{front curvature} in quasi-steady state \textcolor{black}{at the macro scale}. Two mixtures of different regularity were tested, i.e., the highly unstable one of C${_3}$H${_8}$/5O${_2}$ and weakly unstable one of  2C${_2}$H${_2}$/5O${_2}$/21Ar. Firstly, they showed that  detonations in the exponential \textcolor{black}{channels} propagated at a constant average speed, which was controlled by the magnitude of the lateral \textcolor{black}{strain}. Beyond a critical value of \textcolor{black}{lateral strain}, detonations were not possible. Moreover, they compared the experimentally obtained relationship, between the detonation velocity deficit and its front's global curvature, with the generalized ZND model in the presence of lateral \textcolor{black}{strain rate}. The predictions made with the steady ZND model for the velocity deficit disagreed with the experiments. The less unstable 2C${_2}$H${_2}$/5O${_2}$/21Ar detonations showed better agreement between experiments and the theoretical predictions than the more unstable detonations in C${_3}$H${_8}$/5O${_2}$. An interesting question then arises that, can the extended ZND model better predict detonation dynamics of much less unstable mixtures, in spite of the  cellular structures? This becomes the objective of the present work.

It is well known that argon-diluted 2H${_2}$/O${_2}$ detonations have the weakest \textcolor{black}{one dimensional (1D)} instability among those typically investigated experimentally \cite{SHEPHERD200983, radulescu2003propagation} and have a very regular cellular structure. The chemical kinetics of H${_2}$ decomposition is also better known than for hydrocarbons. Therefore, the present study aims to extend the above well-posed technique to more stable mixtures of 2H${_2}$/O${_2}$/2.0Ar, 3.0Ar, 4.5Ar, and 7.0Ar, for the purpose of investigating in detail the dynamics of very regular \textcolor{black}{cellular} detonations \textcolor{black}{with a constant mean lateral strain rate} in exponentially diverging channels. 

\section{Experimental Details}
\label{Experimental Details}

\begin{figure}[]
	\centering
	\subfloat[]{\includegraphics[width=1.0\textwidth]{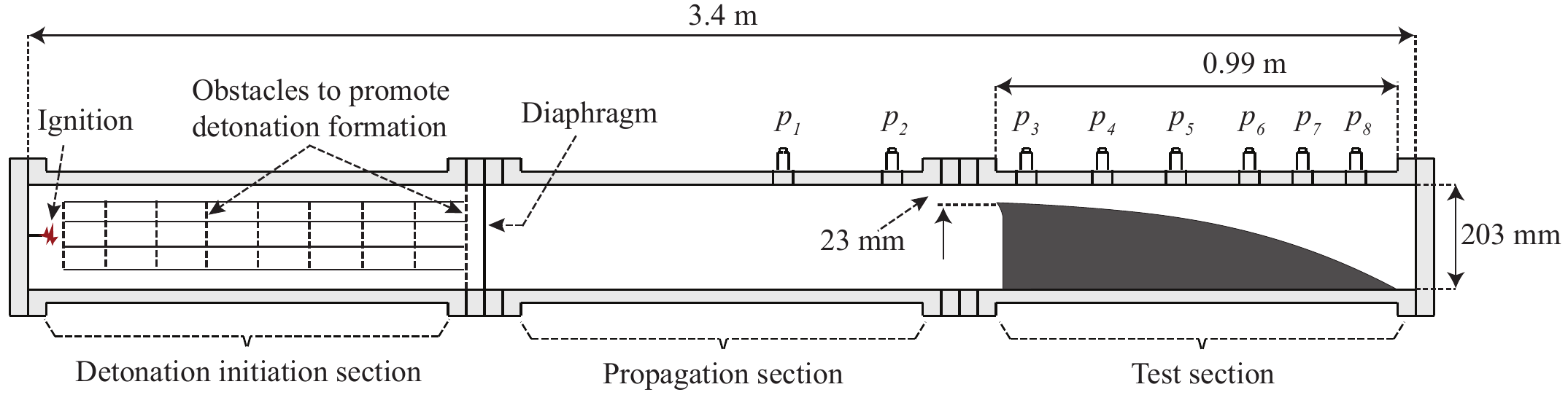}}  \label{shock-tube} \hfill
	\subfloat[]{\includegraphics[width=0.6\textwidth]{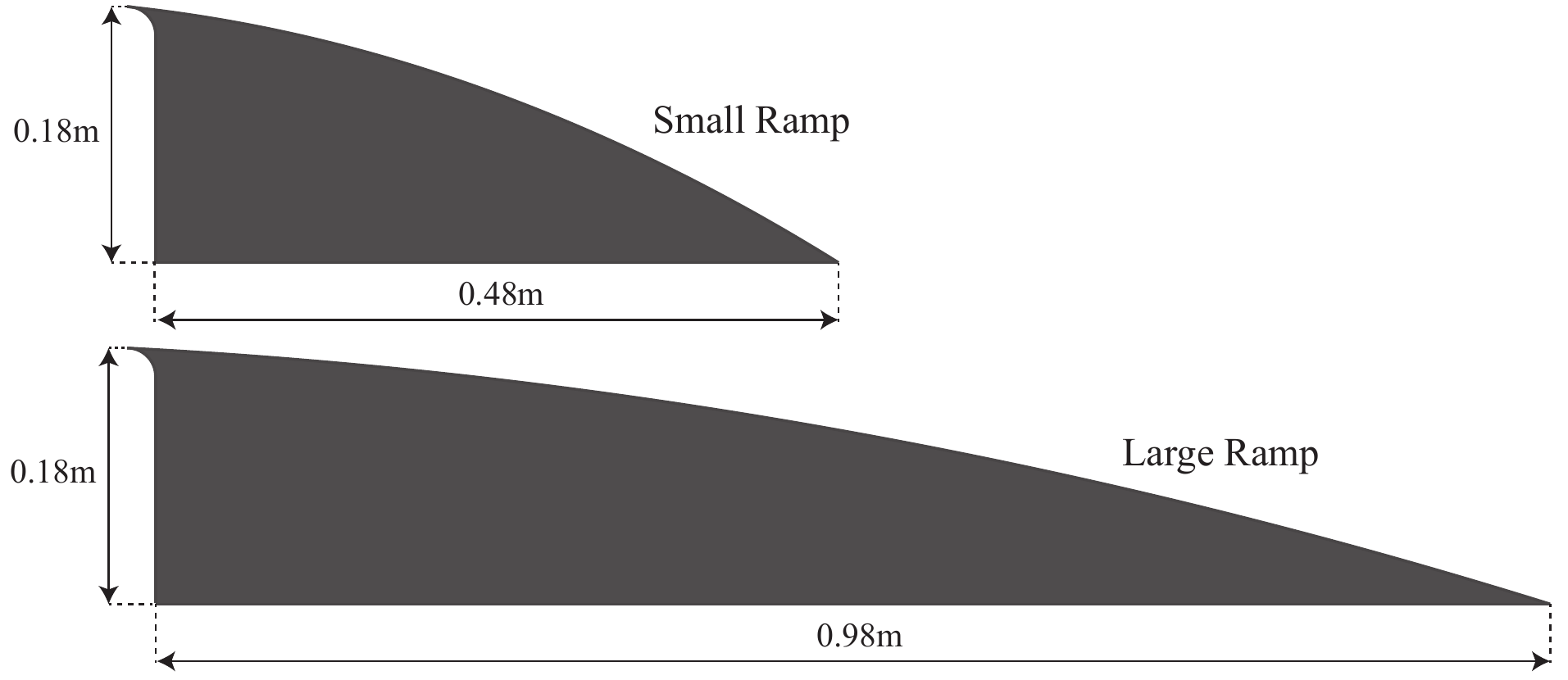}}  \label{ramp}
	\caption{Experimental set-up for diverging detonation experiments: (a) the shock tube with the large ramp inserted in the test section and (b) the exponentially diverging ramps.}
	\label{Experimental-setup}
\end{figure}

The experiments were conducted in a 3.4-m-long aluminium rectangular channel with an internal height and width of 203 mm and 19 mm, respectively. A sketch of the experimental set-up is shown in Fig.\ \ref{Experimental-setup}, which is the same as that adopted by Radulescu and Borzou \cite{Matei-Ramp}. The shock tube comprises three sections, a detonation initiation section, a propagation section, and a test section. The mixture was ignited in the first section by a high voltage igniter (HVI), which could store up to 1000 J with the deposition time of 2 $\mu$s. Mesh wires were inserted in the initiation section for promoting the formation of detonations. Eight high frequency piezoelectric PCB pressure sensors (p1-p8) were mounted flush on the top wall of the shock tube to record pressure signals and then obtain the propagation speeds by using the time-of-arrival method. The test section was equipped with two glass panels in order to visualize the detonation evolution process. For the safety purpose of performing experiments at high initial pressures, the visualization glass panels were alternatively replaced by aluminum ones. 

Two different polyvinyl-chloride (PVC) ramps, which enabled the cross-sectional area $A(x)$ of the channel to diverge exponentially with a constant logarithmic area divergence rate ($K=\frac{d(lnA(x))}{dx}$), were adopted in the test section. Dimensions of the ramps are shown in Fig.\ \ref{Experimental-setup}b. The large ramp had the logarithmic area divergence rate of 2.17 $\textnormal{m}^{-1}$, while for the small one such rate was 4.34 $\textnormal{m}^{-1}$. At the entrance, a protruded rounded tip was kept for minimizing the effects of shock reflection on the detonation front. The initial gap between the ramp tip and the top wall of the channel is 23 mm in height. The height between the exponentially curved wall and the top wall is given by $y_\textrm{{wall}} = y_{0} e^{Kx}$ for $x>0$ and by $y_\textrm{{wall}} = y_{0}$ otherwise.

The mixtures presently studied were stoichiometric hydrogen/oxygen with different argon dilution, i.e., 2H${_2}$/O${_2}$/2.0Ar, 3.0Ar, 4.5Ar, and 7.0Ar.  Each of them was prepared in a separate mixing tank by the method of partial pressures and was then left to mix for more than 24 hours. The mixture was introduced into the shock tube through both ends of the tube at the desired initial pressure with an accuracy of 70 Pa. Before filling with the test mixture in every single experiment, the shock tube was evacuated  below the absolute pressure of 90 Pa. A driver gas of C${_2}$H${_4}$/3O${_2}$ separated by a diaphragm, as shown in Fig.\ \ref{Experimental-setup}a, 
was used in the initiation section for low initial pressures, under which detonations of the test gas cannot be initiated successfully before entering the test section. For visualizing the detonation evolution process along the exponential ramp, a large-scale shadowgraph system was adopted by using a 2m$\times$2m retro-reflective screen with an incandescent filament light source of 1600 W Xenon arc lamp from Newport. The resolution of the high-speed camera was 1152$\times$256 $\textnormal{px}^{2}$ with the frame rate of 42049 fps or 42496 fps. The exposure time was set to 0.81 $\mu$s. Alternatively, a Z-type schlieren setup \cite{settles2001schlieren} with a vertical knife edge was also utilized with a light source of 360 W. The resolution of the high-speed camera was 384$\times$288 $\textnormal{px}^{2}$ with the framing rate of 77481 fps and the exposure time of 0.44 $\mu$s. Note that the background of each shadowgraph and schlieren photograph in the present study was appropriately removed and the images were post-processed.

\section{Experimental Results}
\subsection{Propagation of 2H${_2}$/O$_2$/2Ar detonations along the large ramp}

\begin{figure}[]
	\centering
	{\includegraphics[width=1.0\textwidth]{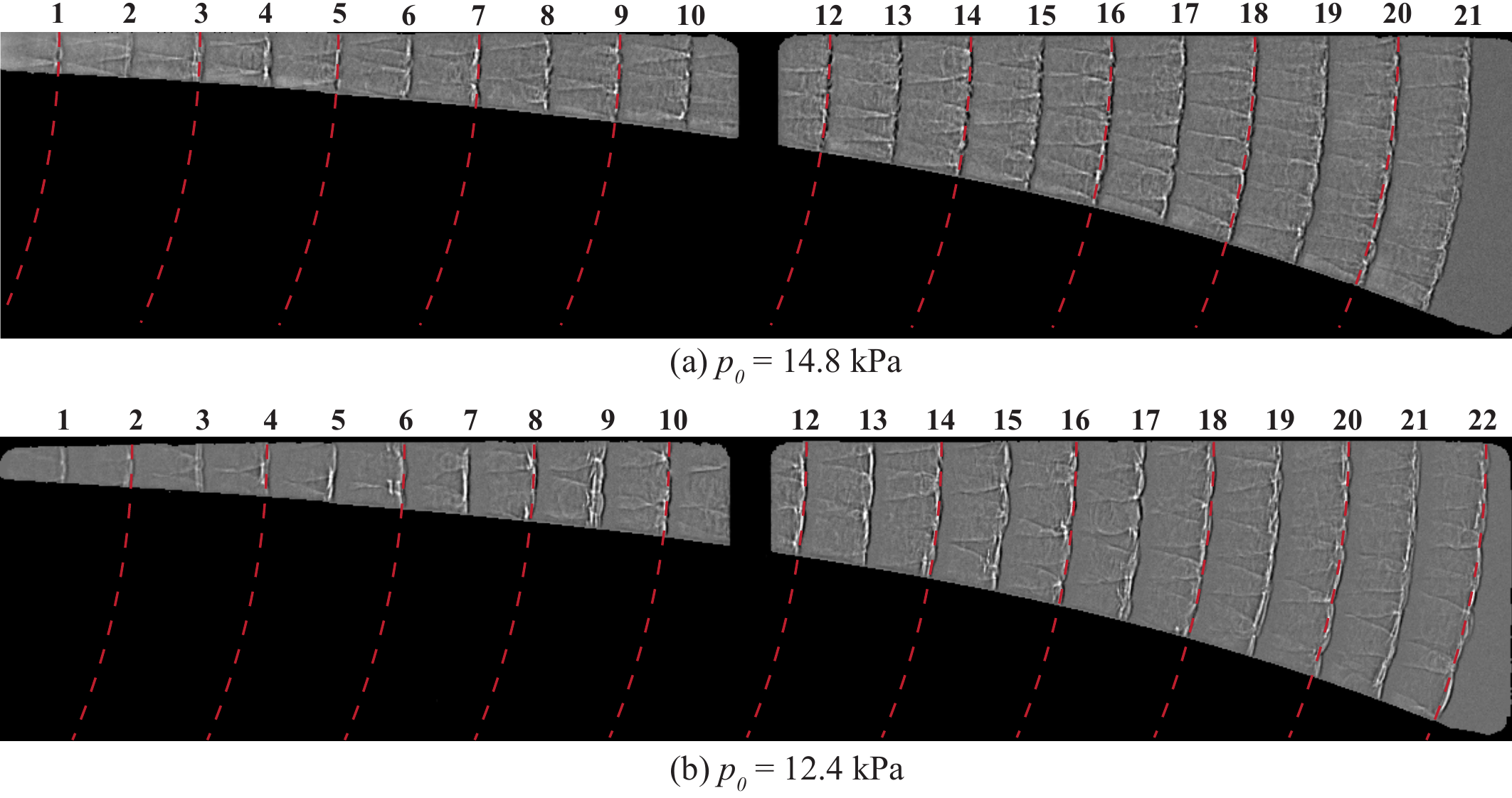}}
	\caption{The superposition of detonation fronts along the large ramp at different instants for the mixture of 2H${_2}$/O$_2$/2Ar; red lines denote arcs of circles with the expected curvature from the quasi-1D approximation.}  \label{2.15psi}  
\end{figure}

The superimposed shadowgraph photos illustrating the evolution of diverging detonation fronts along the large ramp for the mixture of 2H${_2}$/O$_2$/2Ar, at an initial pressure of 14.8 kPa and 12.4 kPa, respectively, are shown in Fig.\ \ref{2.15psi}. The detonation propagated from left towards right. The detonation front acquired a large number of small-sized cellular structures, and was noticeably curved with a characteristic curvature due to the cross-sectional area divergence. Within the limited resolution of the photographs, transverse waves can be recognized starting from triple points and extending backward downstream. One can also note that, as the cross section area of the channel increases, new transverse waves were continuously generated and the average transverse wave spacing appeared to remain constant. This suggests that the cell size remains constant in the self-sustained propagation of diverging detonations along the ramp. The average detonation cell size measured in Fig.\ \ref{2.15psi}a and b is approximately 17.0 mm and 21.0 mm, respectively, which are found to be larger than the values of 10.5 mm and 13.5 mm obtained from the Detonation Database \cite{kaneshige1997detonation} for the same initial pressures.

On the other hand, the theoretically expected arcs of curvature from the quasi-1D approximation, whose radius equals the reciprocal of the logarithmic area divergence rate $K=2.17\textnormal{m}^{-1}$, were obtained and compared with the real detonation fronts in experiments. These arcs of circles with the radius of  $1/K = 0.46 \textnormal{m}$ are denoted by the dashed red lines in Fig.\ \ref{2.15psi}. The comparison shows that the detonation front's global curvature is in very good agreement with that expected by the quasi-1D approximation for designing the exponential geometry, despite some minor deviations near the end of the ramp. \textcolor{black}{These small deviations, as a result of the error in the quasi-1D assumption in designing the exponential geometry, can be evaluated by \cite{Matei-Ramp}
	\begin{equation}
	\frac{K_{2D}}{K} = \left[1+\left(Ky_\textrm{{wall}}\right)^2\right]^{-1/2} \label{deviation-eq}
	\end{equation}
	where the effective curvature $K_{2D}$ represents the curvature of the arc of a circle perpendicularly intersecting both the exponential wall at the point $(x, y_\textrm{{wall}})$ and the top wall as well.}

\begin{figure}[]
	\centering
	{\includegraphics[width=1.0\textwidth]{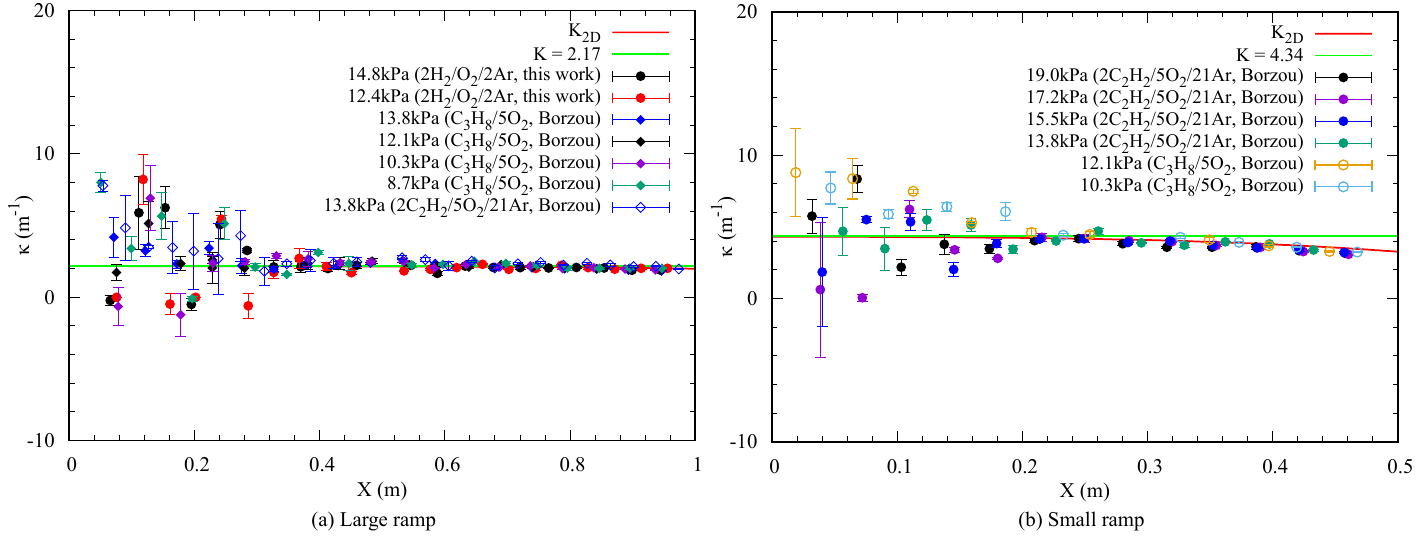}}
	\caption{\textcolor{black}{ The mean curvature of the experimentally obtained curved detonation fronts at different instants. The green line is the constant curvature expected from the quasi-1D approximation while the red line denotes the effective curvature $K_{2D}$ of the geometry as shown in Eq.\ (\ref{deviation-eq}).  }}  \label{KappaSteady}  
\end{figure}

\textcolor{black}{ Figure\ \ref{KappaSteady} shows the evolution of the mean curvature of detonation fronts shown in Fig.\ \ref{2.15psi} as well as for detonations in other mixtures from Borzou's experiments \cite{borzou2016influence}. The front curvature was fitted by using the least-squares method provided by SciPy. Results in Fig.\ \ref{KappaSteady} clearly show that it takes a relaxation length scale for the initially planar detonation front (before entering the ramp) evolving into the curved one due to the exponential geometry. For the large ramp, the relaxation length is approximately 0.4 m, while approximately 0.2 m for the small one. After the relaxation stage, detonations follow the evolution of the effective curvature $K_{2D}$ and can be reasonably assumed to obey a constant mean curvature in their propagation, in spite of the slightly decreasing curvature recognized near the end of the ramps. It thus suggests the appropriateness of the macro-scale quasi-1D approximation for detonations propagating inside these exponential channels.}
 
\begin{figure}[]
	\centering
	{\includegraphics[width=1.0\textwidth]{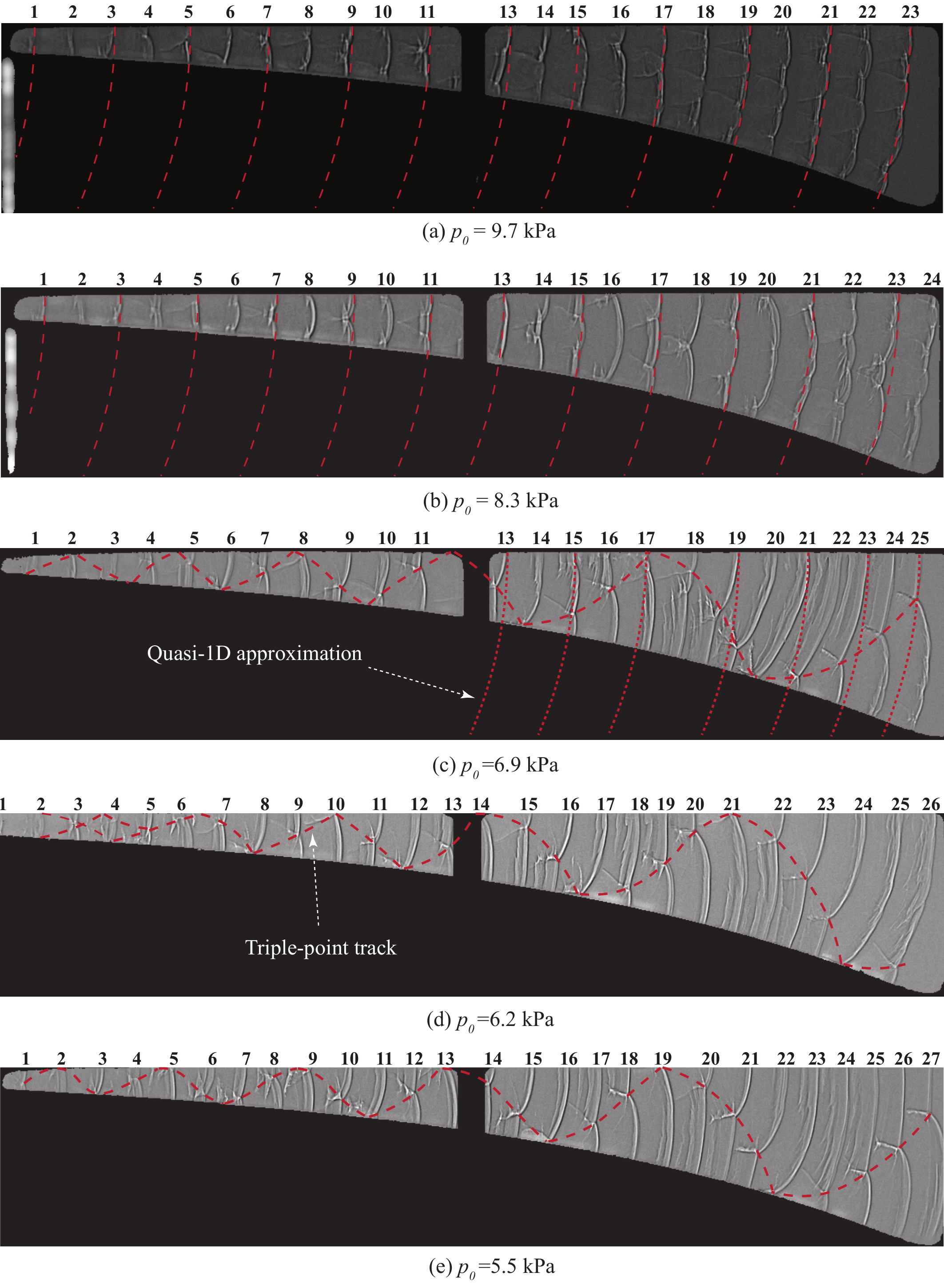}}
	\caption{\textcolor{black}{The evolution of 2H${_2}$/O$_2$/2Ar detonation fronts along the large ramp at different initial pressures near the propagation limit.} } \label{superposition2}  
\end{figure}

\begin{figure}[]
	\centering
	{\includegraphics[width=1.0\textwidth]{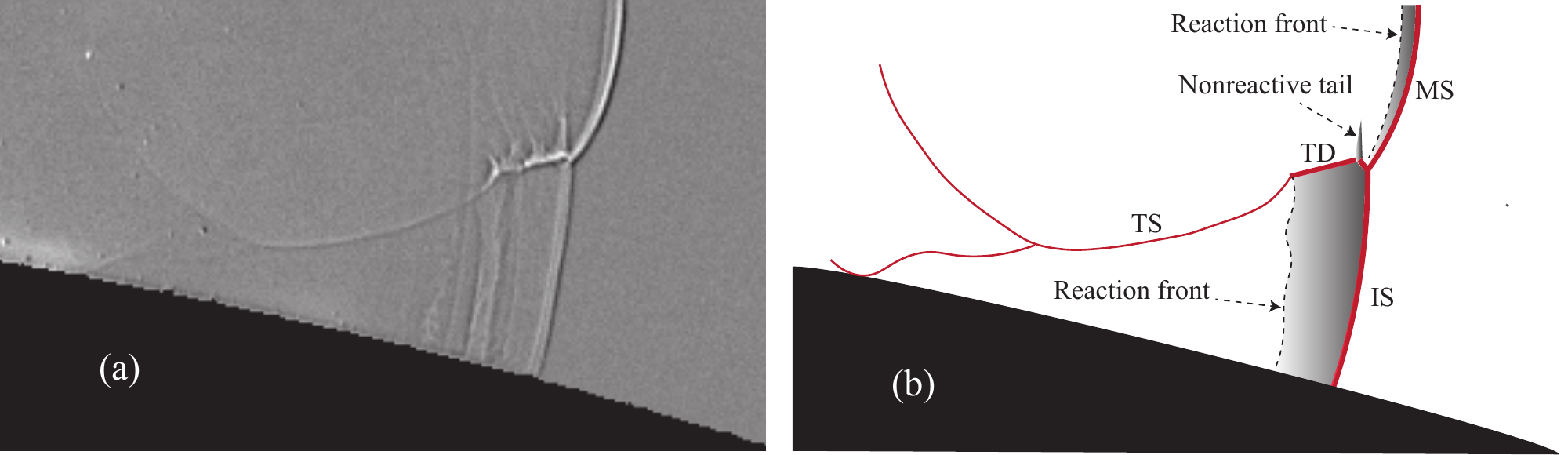}}
	\caption{Details of the characteristic transverse detonation: (a) zoom-in of the cellular structure  of Frame 21 in Fig.\ \ref{superposition2}e, and (b) sketch of the main features. MS: Mach Stem, IS: Incident Shock, TD: Transverse Detonation, TS: Transverse Shock. Note that the shadow zone between the reaction front and the detonation front is the unburned induction zone.} \label{zoom}  
\end{figure}

The detonation propagation process for other lower initial pressures, near the limit, is illustrated in the superimposed shadowgraph photos of Fig.\ \ref{superposition2}. With the decrease of the initial pressure for reducing the kinetic sensitivity of mixtures, detonation cells are considerably enlarged. The triple-shock structure, comprising a Mach stem, an incident shock, and a transverse wave that extends behind the detonation front, can now be clearly observed. One can also observe the consumption of the unreacted induction zones behind the incident shock by the passage of transverse waves, e.g., see Frame 15 through Frame 18 in Fig.\ \ref{superposition2}b, implying the reactivity of these waves. This type of reactive transverse wave has been investigated first in detail by Subbotin \cite{subbotin1975two} for the marginal, weakly unstable detonations and subsequently reported in a series of numerical and experimental works, e.g., Sharpe \cite{sharpe_2001}, Pintgen et al. \cite{PINTGEN2003211}, and Austin \cite{austin2003role}. When the initial pressure was further decreased to approach the failure conditions, below which detonations are not possible, the reaction zone structure \textcolor{black}{becomes very clear} (Fig.\ \ref{superposition2}c-e). These near-limit detonations were able to travel successfully with only one reactive transverse wave, which is interpreted as a transverse detonation \cite{gamezo2000fine}, as can be easily seen from Fig.\ \ref{superposition2}c to Fig.\ \ref{superposition2}e. \textcolor{black}{The main features of the transverse detonation, as shown clearly in Fig.\ \ref{zoom},  are qualitatively similar to that documented numerically by Gamezo et al. \cite{gamezo2000fine}. The transverse detonation, propagating transversely along the large induction zone, is strong enough to burn almost all the material except for a thin tail in the vicinity of the triple point \cite{gamezo2000fine}. The generation mechanism of this thin non-reactive tail in the gap between the leading shock and the transverse detonation front has been clarified as a result of the low temperature of the mixture in an embedded double Mach reflection \cite{gamezo2000fine}.} When delineating the track of the triple point, one can readily obtain the single-headed detonation cell, denoted by the red dashed line in Fig.\ \ref{superposition2}c-e. Clearly, as the detonation propagated towards the end, the detonation cell size considerably increased. An alternative explanation of the enlarging cells is the stabilization mechanism proposed by Short et al. \cite{short2019propagation}. Clearly in these curved detonations, the growth of the detonation front's area may be higher than the intrinsic transverse instability growth rate, thus resulting in the single-headed detonation of continuously increasing cell size without birth of any new transverse waves. \textcolor{black}{At these near limit conditions, significant departures between the real detonation fronts and the arcs of circles of expected curvature in Fig.\ \ref{superposition2}a-c can also be observed.}

\begin{figure}[]
	\centering
	{\includegraphics[width=1.0\textwidth]{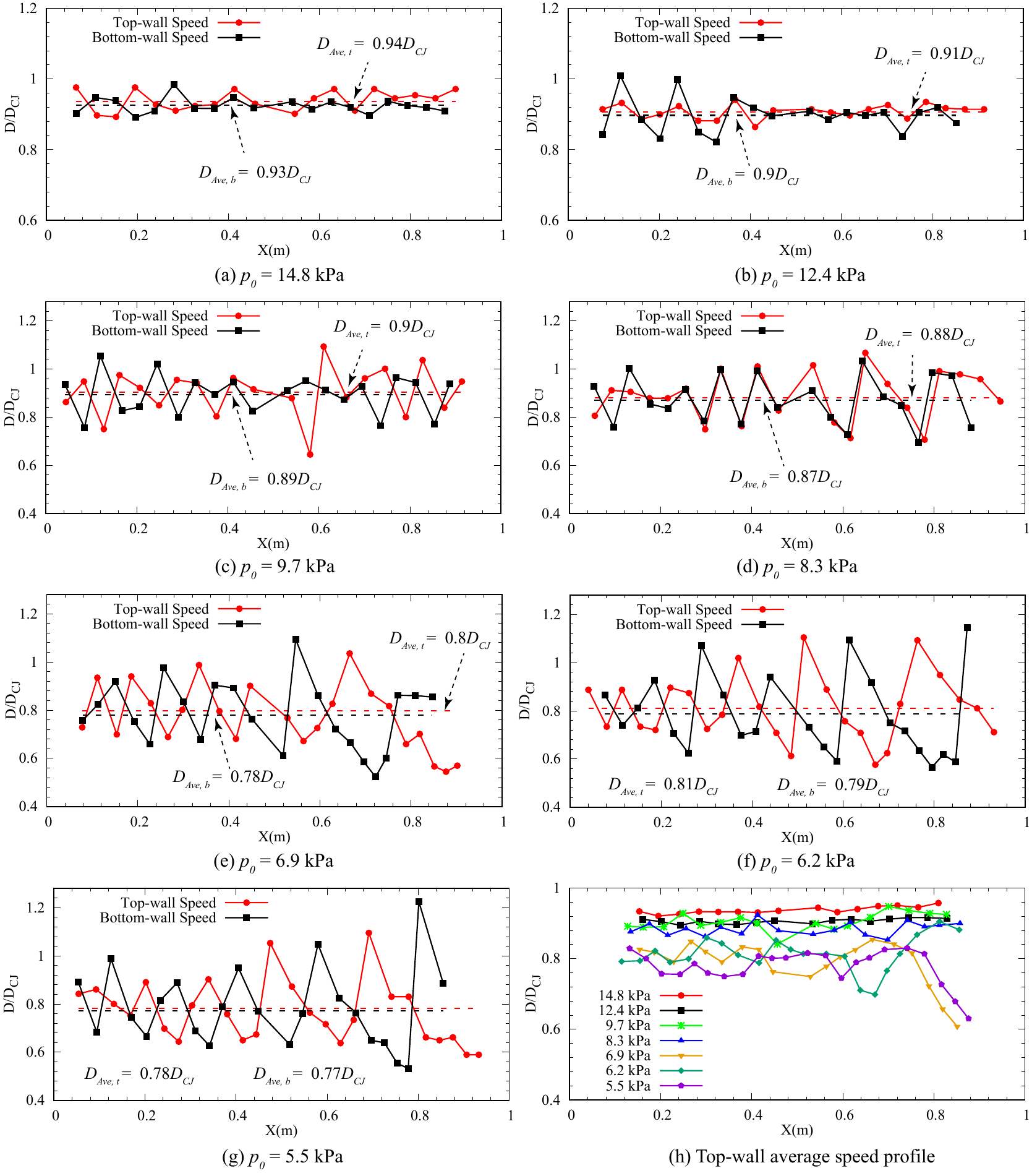}}
	\caption{\textcolor{black}{  The local speed profiles (a)-(g) and the running time average (5$\Delta t$) of the top-wall local speeds (h), of detonations at different initial pressures in Fig.\ \ref{2.15psi} and Fig.\ \ref{superposition2}. The dashed lines represent the corresponding global mean propagation speed.}}  \label{LocalSpeed}  
\end{figure}

\textcolor{black}{The locally averaged speeds both along the top and bottom curved walls were calculated from the  shadowgraph photos in Fig.\ \ref{2.15psi} and Fig.\ \ref{superposition2}, with the distance  between every two neighboring frames divided by their time interval $\Delta t$, and their profiles are shown in Fig.\ \ref{LocalSpeed}.} The speed was normalized by the  ideal CJ detonation velocity calculated with the NASA chemical equilibrium code CEA \cite{mcbride1996computer}. \textcolor{black}{For calculating the speed along the curved wall, the arc length $L_{12}$ of the wall segment between two points $(x_1,y_\textrm{wall$_1$})$ and $(x_2,y_\textrm{wall$_2$})$ can be evaluated by
\begin{equation}
	L_{12} = \frac{1}{K}\left[\sqrt{1+\left(Ky_\textrm{{wall}}\right)^2}+ln\left(\frac{Ky_\textrm{{wall}}}{1+\sqrt{1+\left(Ky_\textrm{{wall}}\right)^2}}\right)\right]^2_{1}
\end{equation}}\textcolor{black}{The results in Fig.\ \ref{LocalSpeed}a-g demonstrate that within the current resolution, the local speed profiles show more significant variations for detonations at lower initial pressures. Especially for the critical detonations with one triple point, as shown in Fig.\ \ref{LocalSpeed}e-g,} the local speed profiles exhibit periodic fluctuations, ranging from the maximum of 1.2 $D_{CJ}$ to the minimum of 0.5 $D_{CJ}$. These characteristic fluctuations very well illustrate the \textcolor{black}{periodic evolution of the lead shock inside the cells}. As the detonation propagated to the right end, the fluctuation period increased, implying an increasing length scale of the detonation cell. This is consistent with the finding from Fig.\ \ref{superposition2} that the detonation cell size increased as a result of the diverging area for the near-limit detonations. \textcolor{black}{Noteworthy are the out-of-phase periodic velocity profiles measured along the top and bottom walls, shown in Fig.\ \ref{LocalSpeed}e-g. This can be interpreted as a result of the alternation between the strong Mach stem with relatively high speed and the weak incident shock of low velocity in the detonation front of a single triple point. The global mean propagation speeds along the top wall and bottom walls differed by at most 3$\%$ of $D_{CJ}$. Moreover, the running time average (of $5\Delta t$) of the top-wall local speeds has also been shown in Fig.\ \ref{LocalSpeed}h. It again illustrates that for initial pressures well above the limit, detonations can be assumed to propagate in quasi-steady state at a macro scale much larger than the cellular structure with a constant speed (e.g., see the profiles of 14.8 kPa and 12.4 kPa).}   

\begin{figure}[]
	\centering
	{\includegraphics[width=1.0\textwidth]{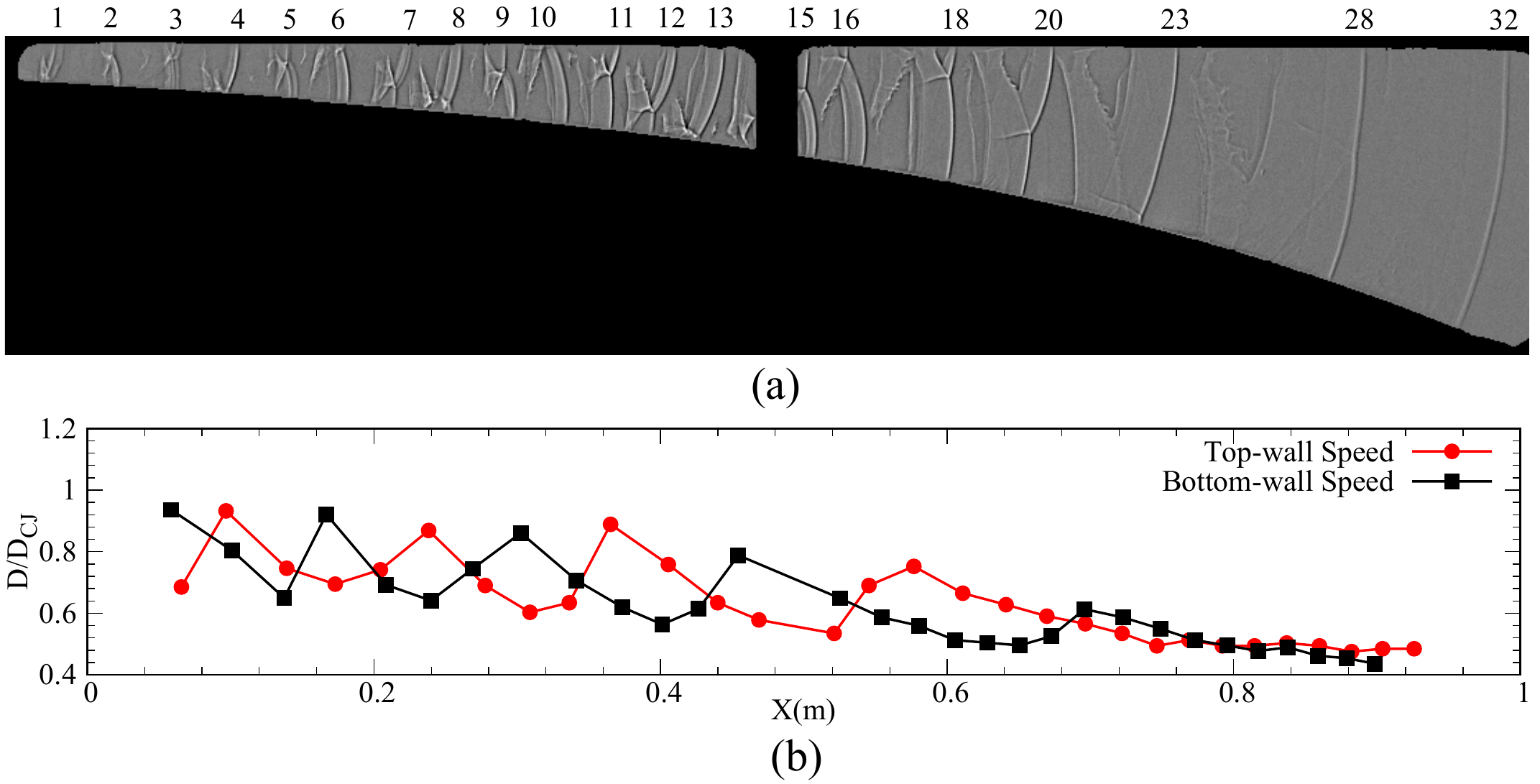}}
	\caption{Shadowgraph photos (a) and local speed profiles (b) of a failed detonation at the initial pressure of 5.5 kPa for the mixture of  2H${_2}$/O$_2$/2Ar.} \label{failure}  
\end{figure}

For the critical pressure of 5.5 kPa, below which detonations were unable to propagate, six experiments in total were repeated. It was found that three of them \textcolor{black}{successfully propagated as single-headed detonations} while the rest three finally failed. \textcolor{black}{This apparently stochastic behavior near the limits has also been observed in} studies on detonation diffraction \cite{Higgins-propane, mevelhydrogen}. Any slightly different perturbations during the whole evolution process can sensitively impact the result of Go or No-Go for detonation propagation in the critical pressure range. Figure\ \ref{failure} shows the failing process of the critical detonation and correspondingly its speed profiles. Before the failure, it was similar to that of the successful cases in the propagation mechanism of a single-headed detonation. As the detonation proceeded, the trailing reaction front was gradually detached from the leading shock and unreacted tongues were formed behind, with the transverse wave becoming inert. Finally, the detonation failed with the complete detachment of the reaction front from the leading shock (see Frame 28 in Fig.\ \ref{failure}a), thereby resulting in the continuous decay of the shock, which can be seen from the continuously decreasing speed in Fig.\ \ref{failure}b. Of noteworthy is that no distinctive transverse detonations were observed in the decoupled shock-flame complex for the failure case. 

\begin{figure}[]
	\centering
	{\includegraphics[width=0.6\textwidth]{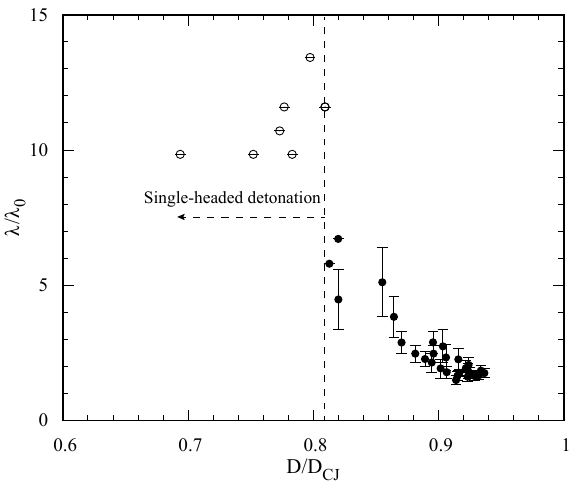}}
	\caption{The ratio of estimated cell size ($\lambda$) in the present large ramp experiments to previously reported cell size ($\lambda_0$) as a function of the normalized mean propagation speed $D/D_{CJ}$ for detonations in the mixture of  2H${_2}$/O$_2$/2Ar.  } \label{Cellsize}  
\end{figure}

In addition, the average detonation cell size was also obtained from the shadowgraph photos of each experiment, as illustrated in Fig.\ \ref{Cellsize}. It characterizes the relationship between the ratio of the presently estimated cell size ($\lambda$) to previously reported cell size ($\lambda_0$) \cite{engel1969transverse, barthel1974predicted}, and the mean propagation speed, normalized by the ideal CJ value. Note that the relation of $\lambda_0$ was given by $\lambda_0(p) = 443.3p^{-1.39}$, which was fitted from the data in the Detonation Database \cite{kaneshige1997detonation}. Also, it is reasonable to assume these reported cell sizes for detonations with very limited or no velocity deficits, i.e., ideal CJ detonations. The results from Fig.\ \ref{Cellsize} show that the normalized cell size ($\lambda/\lambda_0$) increases considerably when decreasing the mean propagation speed with respect to the ideal CJ speed ($D/D_{CJ}$), suggesting a strong dependence of the cell size on the detonation velocity deficit. \textcolor{black}{This agrees with the finding of Ishii and Monwar \cite{ishii2011detonation} for detonations in narrow channels of varied sizes}. When the mean propagation speed further decreased to 0.8$D_{CJ}$, detonations started to be organized in a single-headed one, whose cell size was about 10$\sim$15 times larger than that of the ideal CJ detonation at the same initial pressure. Previous works proposed the onset of single-headed spinning detonations as a criterion for propagation limits of detonations in narrow tubes \cite{Moen-1081, Lee-2013}. Here, it further demonstrates the characterization of propagation limits inside the exponentially diverging channel by a single-headed detonation, organized with a distinctive transverse detonation.

\subsection{Propagation of 2H${_2}$/O$_2$/2Ar detonations along the small ramp}

\begin{figure}[]
	\centering
	{\includegraphics[width=1.0\textwidth]{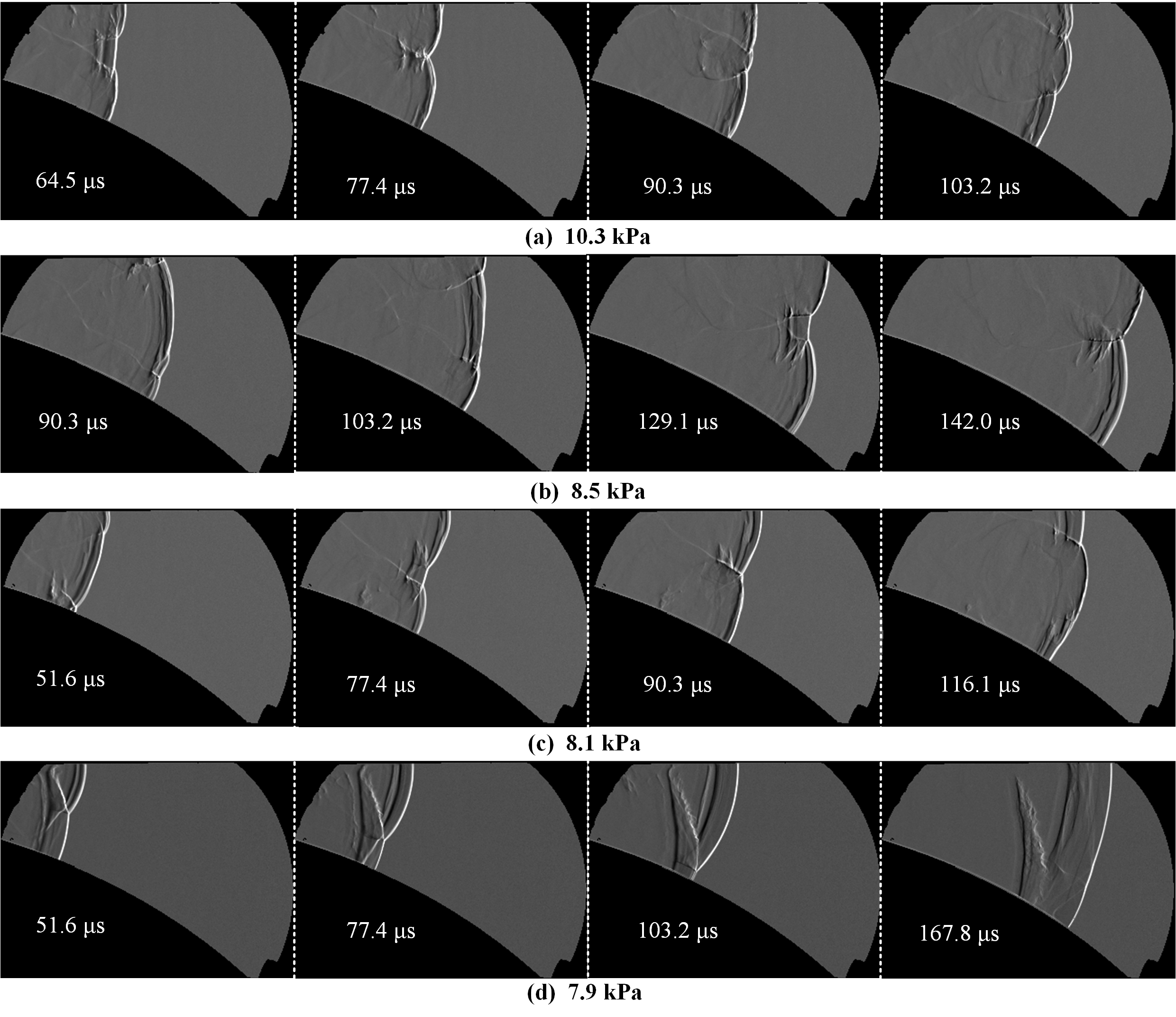}}
	\caption{Schlieren photographs of near-limit detonations along the small ramp for the mixture of  2H${_2}$/O$_2$/2Ar.} \label{SmallRamp}  
\end{figure}

Experiments were also performed for detonations propagating along the small ramp of a higher area divergence rate of $K=4.34\textnormal{m}^{-1}$, i.e., double that of the large ramp.  Visualization of the reaction zone structures of detonations at varied initial pressures near the limit, as shown in Fig.\ \ref{SmallRamp}, enables us to make the following observations. At 10.3 kPa, detonations propagated with two triple points, i.e., one pair of transverse waves, along the detonation front (Fig.\ \ref{SmallRamp}a). Evidently, these transverse shocks were reactive in burning the unburned gases behind the leading shock, preventing the generation of significant unreacted gas pockets \cite{austin2003role, PINTGEN2003211}. When the initial pressure was decreased to approach the critical one of 8.1 kPa, a single-headed detonation featuring a transverse detonation was formed, as is clearly illustrated in Fig.\ \ref{SmallRamp}b and c. Below this critical pressure, the leading shock was found to continuously detach from the trailing reaction front with the transverse wave being inert, which can be seen from Fig.\ \ref{SmallRamp}d. Since the average speeds along the walls were measured to be only 0.5$\sim$0.6 $D_{CJ}$, detonations can thus be interpreted as finally failed at the initial pressure of 7.9 kPa in Fig.\ \ref{SmallRamp}d. \textcolor{black}{It is not clear if such case along the small ramp with a much longer extended length in the stream-wise direction would have been possible. }

\begin{figure}[]
	\centering
	{\includegraphics[width=0.55\textwidth]{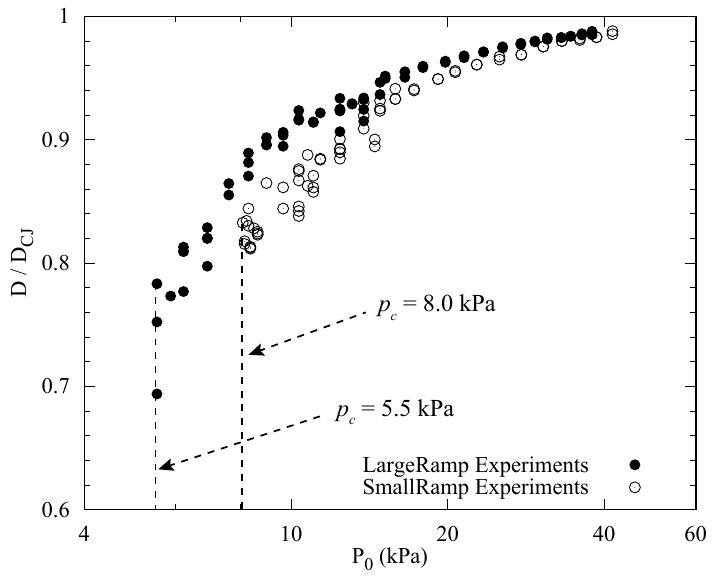}}
	\caption{\textcolor{black}{Average speeds normalized by the corresponding CJ speeds with respect to different initial pressures for the mixture of 2H${_2}$/O${_2}$/2.0Ar.}  } \label{2H2-O2-2Ar-DP}  
\end{figure}

The global mean propagation speeds of 2H${_2}$/O$_2$/2Ar detonations, measured \textcolor{black}{along the top wall} in all the experiments of both ramps, are shown in Fig.\ \ref{2H2-O2-2Ar-DP} as a function of initial pressures. As a result of the flow divergence, experienced by detonations inside the cross-section area diverging channels, the mean propagation speed is smaller than its ideal CJ detonation speed by a velocity deficit. These velocity deficits increase with the rate of geometrical area divergence, as can be easily concluded in Fig.\ \ref{2H2-O2-2Ar-DP} from the higher propagation speeds measured in the large ramp  experiments ($K=2.17\textnormal{m}^{-1}$) than that of the small ramp ($K=4.34\textnormal{m}^{-1}$) under the same initial pressures. As the initial pressure is reduced, the deviation of the mean propagation speed from the ideal CJ value becomes larger, implying a more significant role of area divergence in impacting detonations of lower initial pressures. Near the limit, such velocity deficits can reach 20$\%$ $\sim$ 30$\%$ of the CJ value.  One more observation is the higher propagation limit pressure $p_c$ of detonations along the small ramp. This can be interpreted as the consequence of more losses experienced by detonations along the small ramp due to its larger \textcolor{black}{lateral strain rate}, thus giving rise to a higher critical pressure.

\subsection{\textcolor{black}{Effect of argon dilution}}

\begin{figure}[]
	\centering
	{\includegraphics[width=1.0\textwidth]{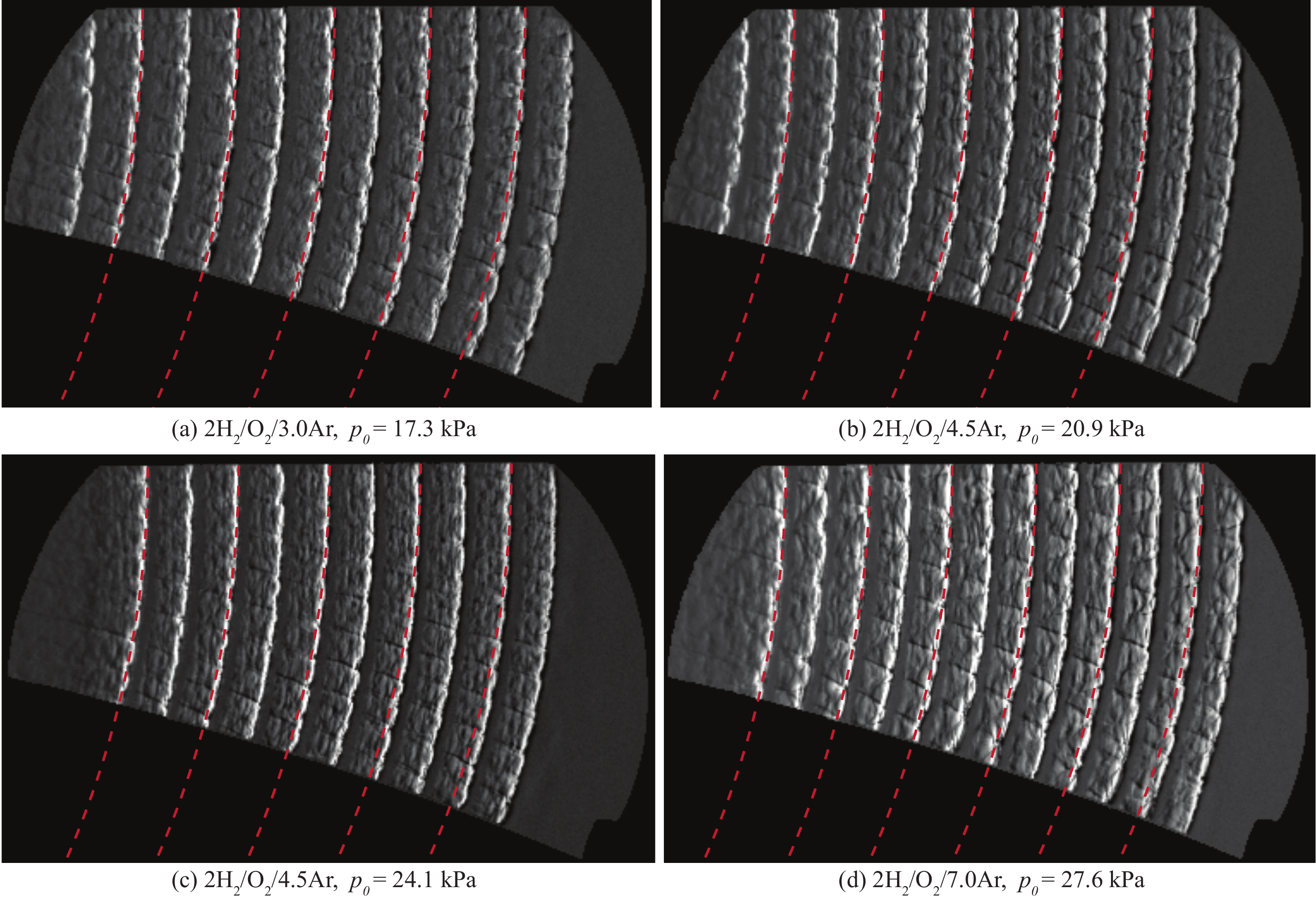}}
	\caption{Comparison of the superimposed detonation fronts near the end of the large ramp with the arcs of curvature (denoted by red lines) expected from the quasi-1D approximation for different mixtures.} \label{largeramp-curvature-fit}  
\end{figure} 

\begin{figure}[!htb]
	\centering
	{\includegraphics[width=1.0\textwidth]{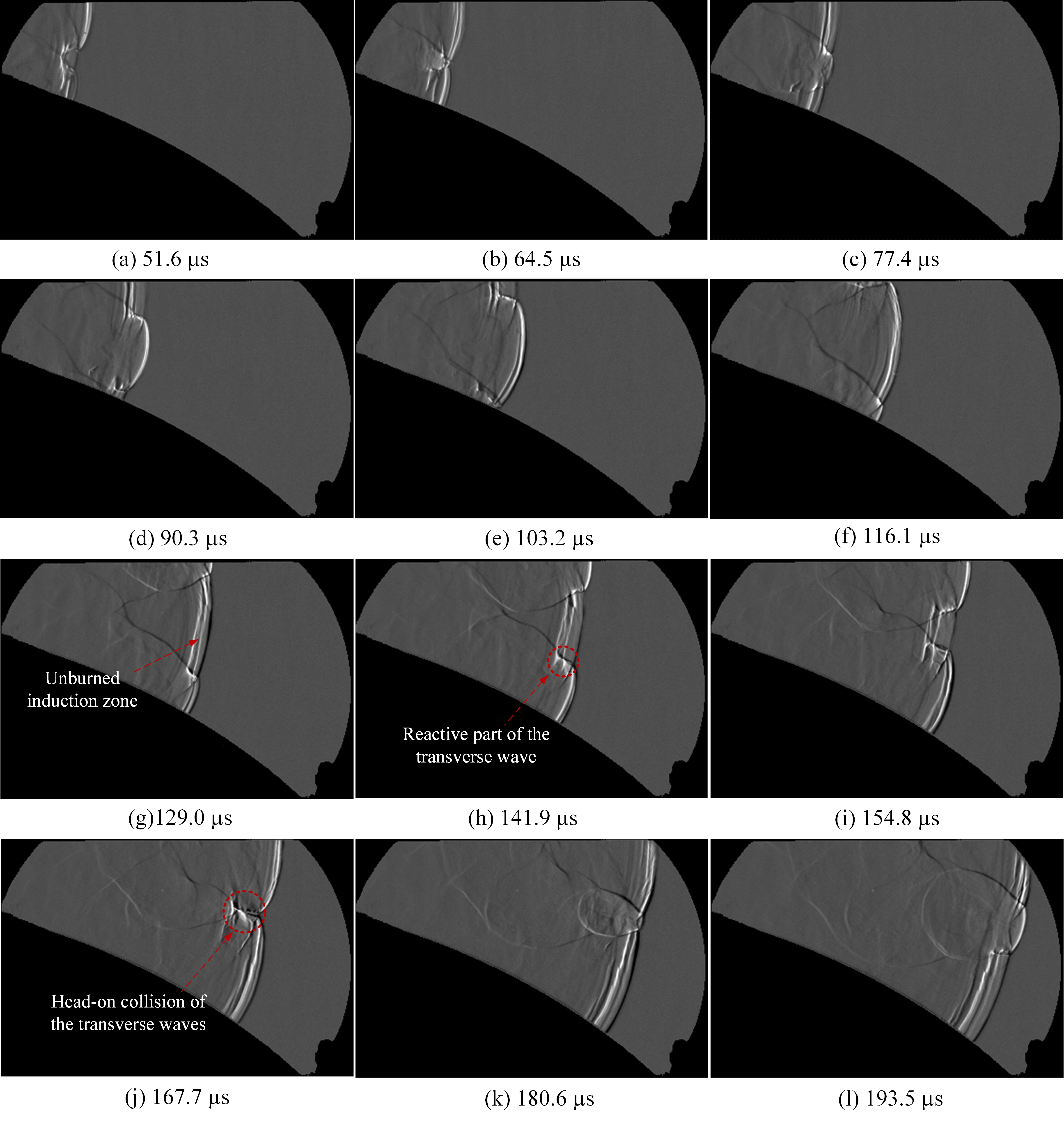}}
	\caption{Schlieren photographs illustrating detonation evolution along the small ramp in the mixture of 2H${_2}$/O$_2$/4.5Ar at the initial pressure of 11.2 kPa.} \label{2H2-O2-4.5Ar-SmallRamp}  
\end{figure}
In the present study, experiments of detonations in stoichiometric H${_2}$/O$_2$ mixtures with other argon dilutions both along the large ramp and small ramp were also performed. It served the purpose of both qualitatively and quantitatively demonstrating the influence of argon dilution on detonation behaviors. The mixtures involved in this part are 2H${_2}$/O$_2$/3.0Ar, 4.5Ar, and 7.0Ar, corresponding to the argon dilution of 50$\%$, 60$\%$, and 70$\%$, respectively. Figure\ \ref{largeramp-curvature-fit} shows the superimposed schlieren photographs illustrating the evolution of detonations, well above the limit, along the large ramp for different mixtures. The curved detonation fronts were uniformly textured with triple points, with transverse waves extending downstream behind the leading shock. Again, cell sizes remain approximately constant in the exponentially diverging channel, under initial pressures far away from the limit. Moreover, comparisons of these experimentally obtained detonation fronts with arcs of the expected curvature from the quasi-1D approximation were also made. The very good agreement between the real curved detonation fronts and the theoretically expected arcs, denoted by the red dashed lines in Fig.\ \ref{largeramp-curvature-fit}, demonstrates the independence of the detonation front's global curvature on mixture compositions and initial pressures. It is thus indicative of the fact that argon diluted H${_2}$/O$_2$ detonations, well above the limit, propagate in quasi-steady state at the macro-scale with a constant mean curvature.

\begin{figure}[!htb]
	\centering
	{\includegraphics[width=1.0\textwidth]{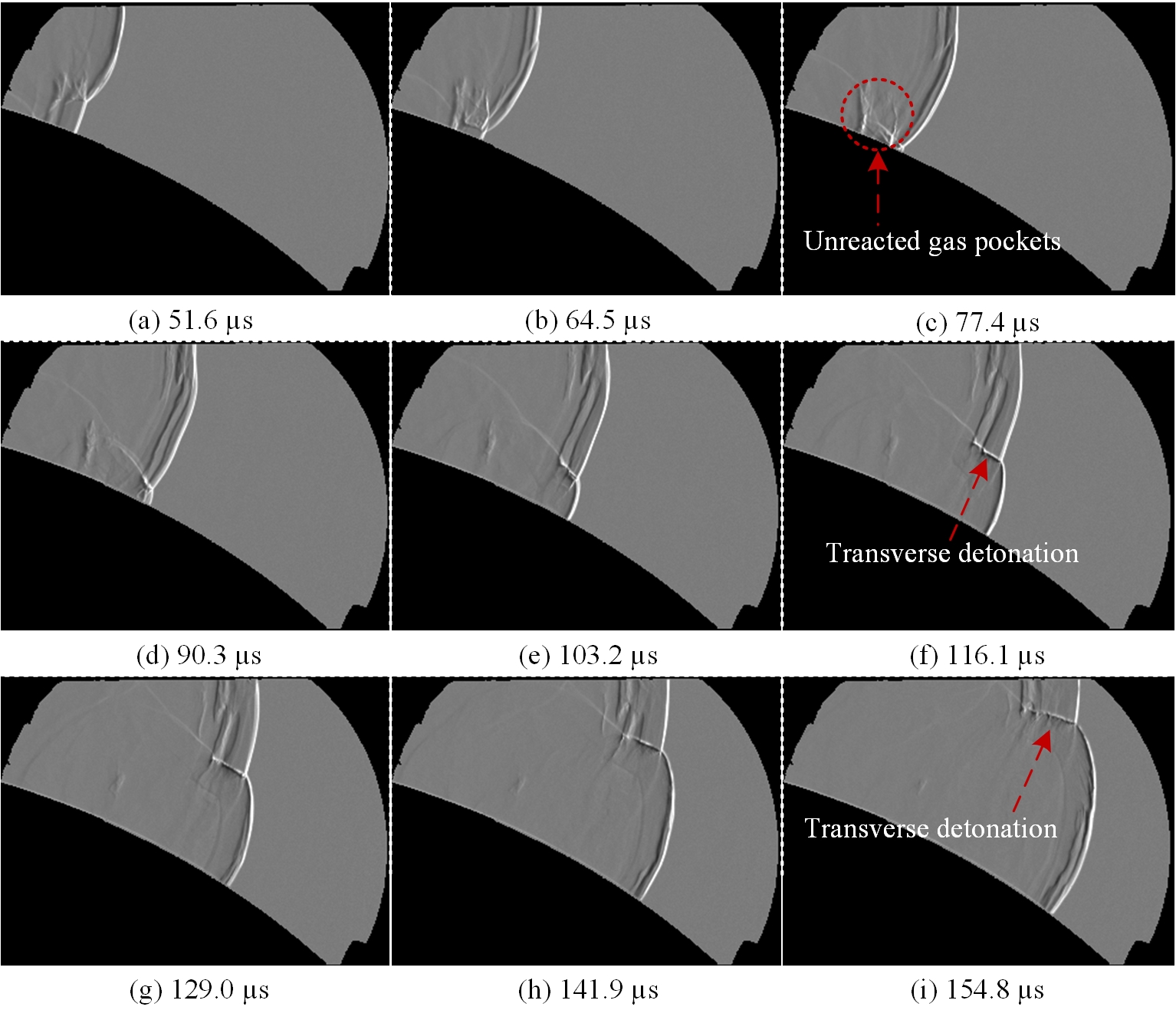}}
	\caption{Schlieren photographs illustrating detonation evolution along the small ramp in the mixture of 2H${_2}$/O$_2$/3Ar at the initial pressure of 8.8 kPa.} \label{2H2-O2-3Ar-SmallRamp}  
\end{figure}

\begin{figure}[!htb]
	\centering
	{\includegraphics[width=1.0\textwidth]{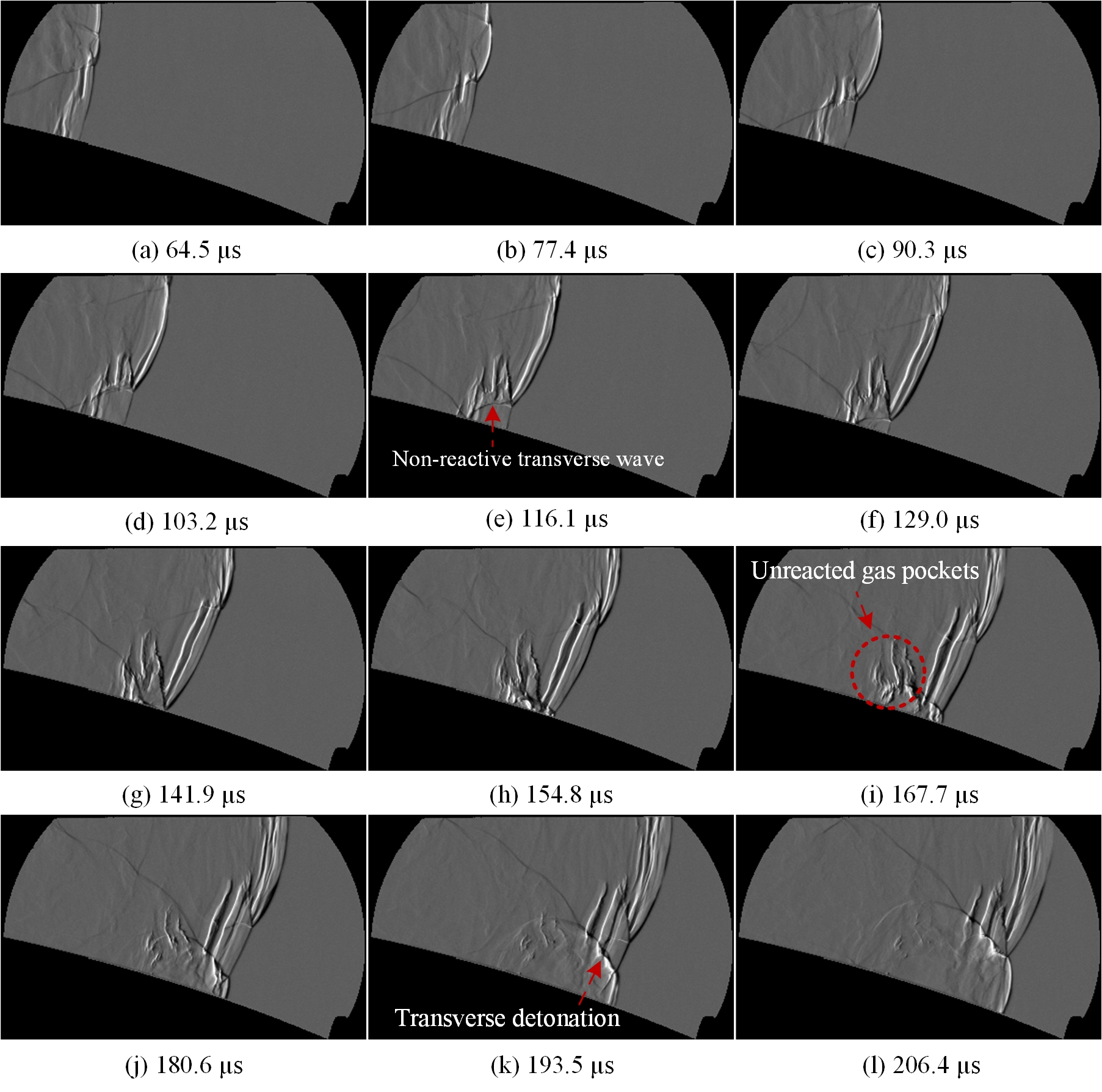}}
	\caption{Schlieren photographs illustrating detonation evolution along the large ramp in the mixture of 2H${_2}$/O$_2$/7Ar at the initial pressure of 13.8 kPa.} \label{2H2-O2-7Ar-LargeRamp-2psi}  
\end{figure} 

The visualized cellular structures of near-limit detonations in mixtures of 2H${_2}$/O$_2$/3.0Ar, 4.5Ar, and 7.0Ar show qualitatively the same behaviors as that of 2H${_2}$/O$_2$/2.0Ar detonations. For example, Fig.\ \ref{2H2-O2-4.5Ar-SmallRamp} gives the evolution process of detonations with one pair of triple-shock structures in the mixture of 2H${_2}$/O$_2$/4.5Ar at the initial pressure of 11.2 kPa. It excellently demonstrates the interactions between the very regular cellular structures, including the collision of transverse waves and their reflections from the walls. Due to the reactive portion of the transverse shock, sweeping across the unburned induction zone behind the incident shock, no significant unreacted gas pockets were observed. The detailed propagation process of single-headed detonations is illustrated in Fig.\ \ref{2H2-O2-3Ar-SmallRamp} for the mixture of 2H${_2}$/O$_2$/3.0Ar. Prior to the formation of a transverse detonation, some unburned gases can be observed behind the transverse wave in the first three frames, implying that this transverse shock was non-reactive and of weak type. The transverse detonation occurred after the reflection of the inert transverse wave from the curved wall, which can be seen from the frames of 90.3 $\mu$s through 116.1 $\mu$s in Fig.\ \ref{2H2-O2-3Ar-SmallRamp}. This behavior can be better observed in Fig.\ \ref{2H2-O2-7Ar-LargeRamp-2psi}, showing the transition process between the non-reactive and reactive transverse waves in near-limit detonations in the mixture of 2H${_2}$/O$_2$/7.0Ar. The initially reactive transverse wave decayed gradually and transitioned to a non-reactive transverse wave, which could be seen from the frames of 64.5 $\mu $s through 141.9 $\mu$s (Fig.\ \ref{2H2-O2-7Ar-LargeRamp-2psi}). After the reflection of the non-reactive transverse wave from the bottom curved wall, considerable unreacted gases were pinched off as pockets, e.g., see frames of 154.8 $\mu$s and 167.7 $\mu$s. The reflected transverse wave again became reactive and rapidly transformed to a strong transverse detonation burning all of the unreacted gases behind the leading shock. After the generation of the transverse detonation, unreacted gas pockets were no longer produced. These observations confirm Subbotin's finding of both reactive and non-reactive transverse waves existing in marginal detonations \cite{subbotin1975two} and the mechanism of forming unreacted gas pockets due to inert transverse waves \cite{subbotin1975two,gamezo2000fine,sharpe_2001,austin2003role}.

\begin{figure}[!htb]
	\centering
	{\includegraphics[width=1.0\textwidth]{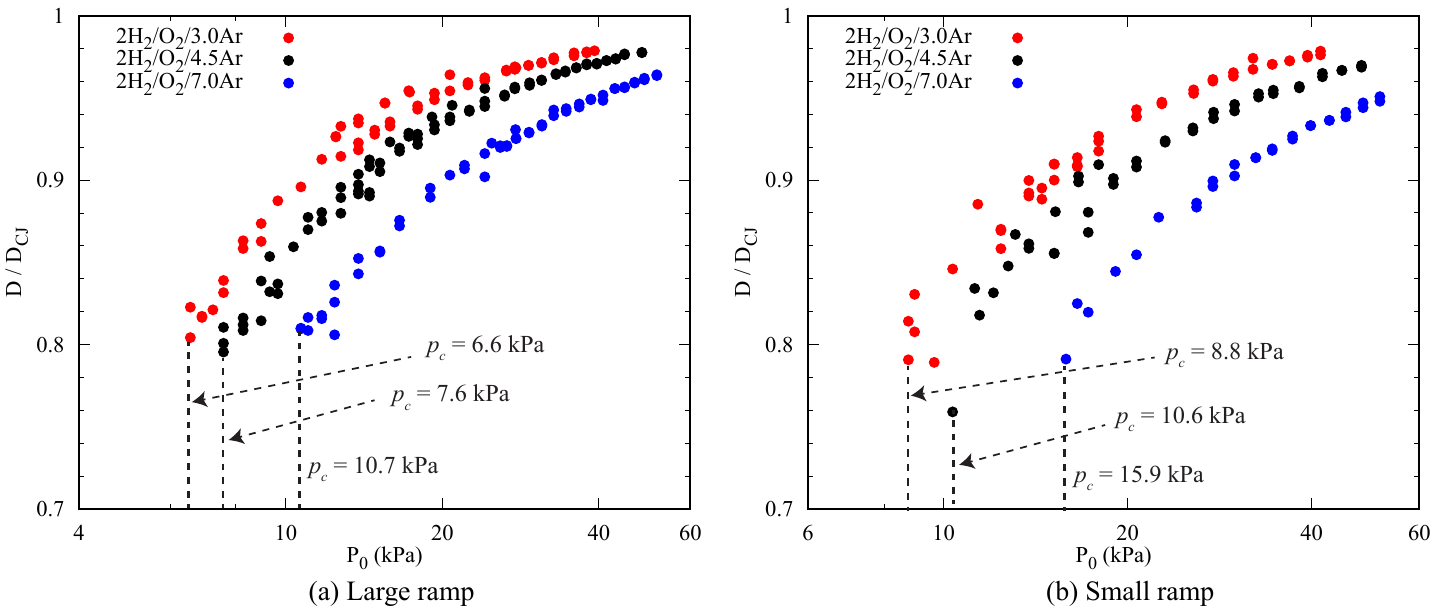}}
	\caption{Global mean propagation speeds of (a) along the large ramp and (b) along the small ramp with respect to initial pressures for different mixtures. } \label{Pressure-Velocity-ArDiluetion}  
\end{figure}

The relationships between the mean propagation speeds and initial pressures are shown in Fig.\ \ref{Pressure-Velocity-ArDiluetion} for detonations in these mixtures, along the two ramps. As the argon dilution increases, detonations of the same initial conditions propagate with a larger velocity deficit, and the critical pressure $p_c$ also increases. This can be interpreted in terms of the gas sensitivity varying with the increase in dilutions of argon.  Mixtures with higher argon dilutions are the ones with reduced reaction rates and chemical energy release rates, thus giving rise to larger velocity deficits and limiting pressures. Near the limit, the measured velocity deficits can reach 20$\%$ $\sim$ 25$\%$ of the CJ value, as is consistent with that of 2H${_2}$/O$_2$/2Ar.  

\section{Discussion}
\subsection{\textcolor{black}{Relaxation to quasi-steady state}}

\textcolor{black}{The investigation of the entrance effects in our experiments shown in Fig.\ \ref{KappaSteady} suggest that the relaxation length scale for the wave to adopt a steady state was approximately on the order of the inverse steady state curvature (i.e., the radius of curvature $1/K$ ), i.e., $1/2.17 = 0.46$ m and $1/4.34 = 0.23$ m, respectively for the two ramps investigated. }

\begin{figure}[!htb]
	\centering
	{\includegraphics[width=0.69\textwidth]{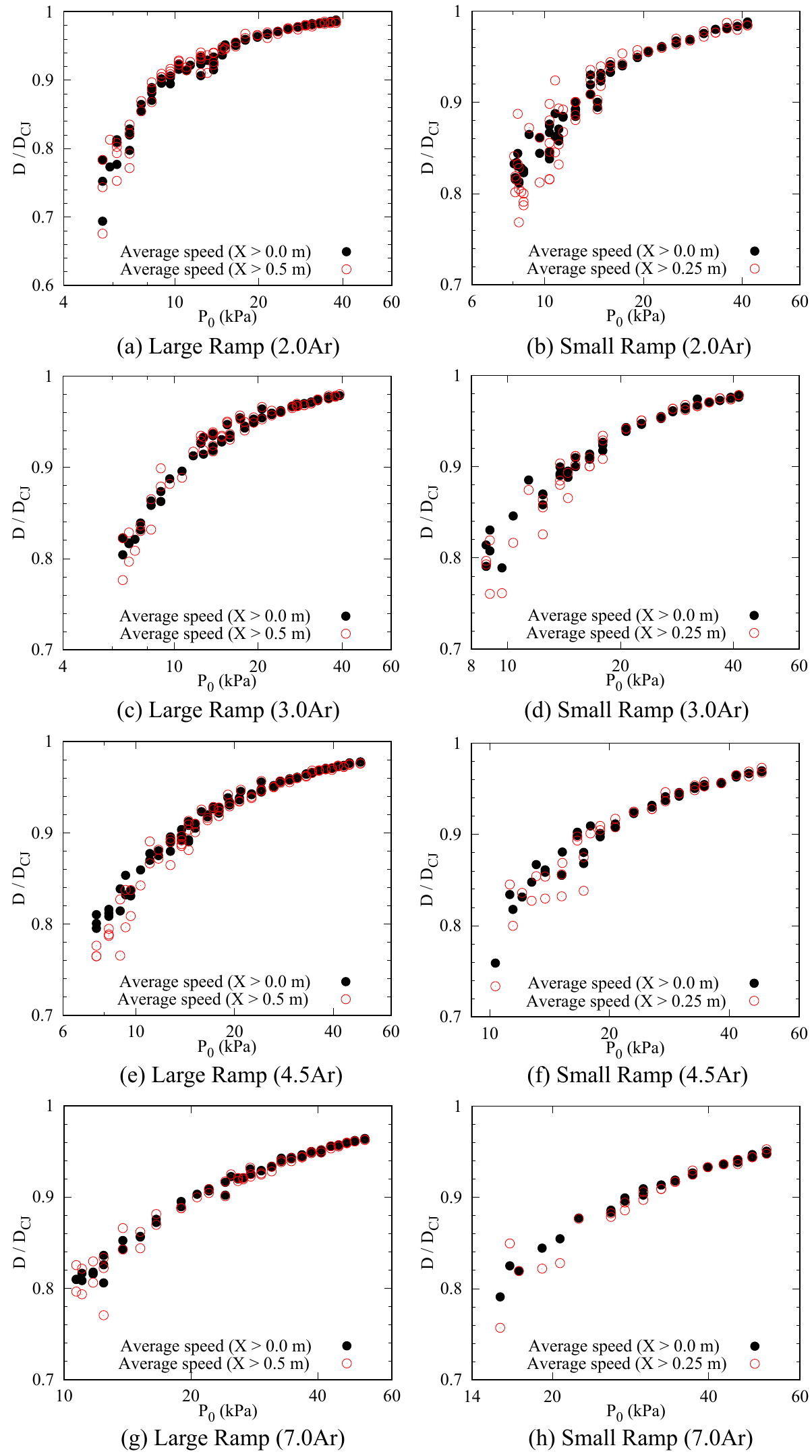}}
	\caption{Comparison of the global mean propagation speed ($X > 0.0$ m) with the average speed over the latter half section of the ramp ($X > 0.5$ m for the large one, while $X > 0.25$ m for the small one).  } \label{Pressure-Velocity-ArDiluetion-bothaverage}  
\end{figure}

\textcolor{black}{The relative non-importance of the entrance effects is also borne out from the comparison of the detonation speeds obtained in the first and second half of the channel, shown in Fig.\ \ref{Pressure-Velocity-ArDiluetion-bothaverage} for all the experiments performed.  The very good consistency between the mean propagation speeds averaged over the whole section and those over the latter half section further confirms the appropriateness of assuming the macro-scale quasi-steady detonations inside the exponential channels.}

\begin{figure}[!htb]
	\centering
	\subfloat[]{\includegraphics[width=1.0\textwidth]{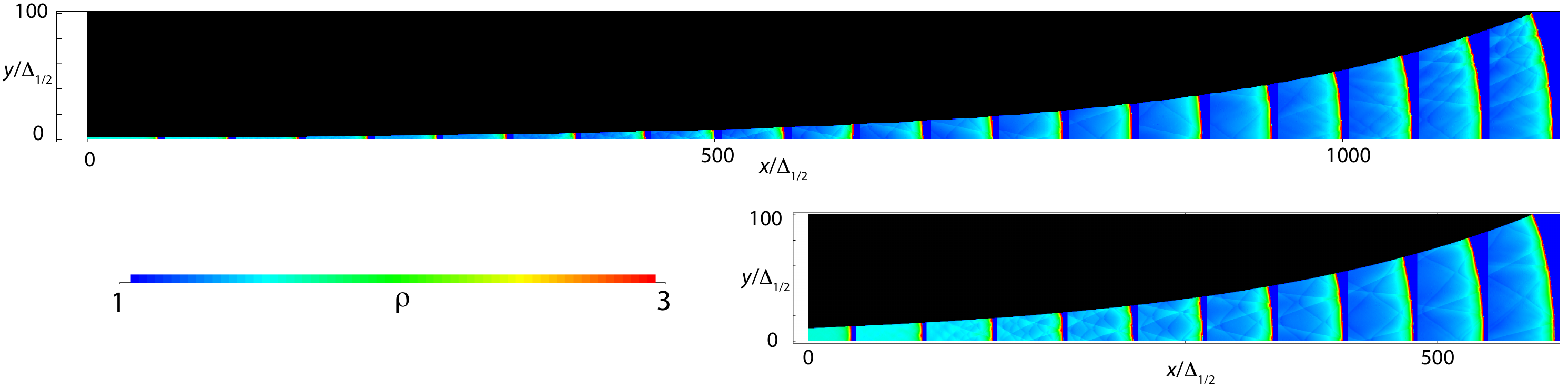}}  \label{fronts-evolution} \hfill
	\subfloat[]{\includegraphics[width=0.49\textwidth]{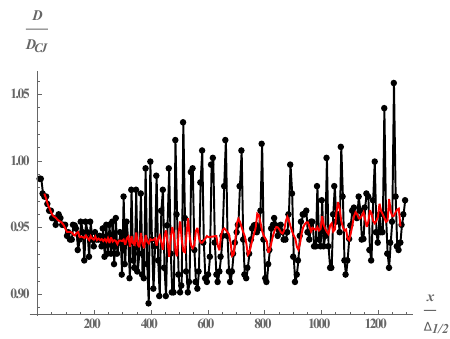}}  \label{speeds1}
	\subfloat[]{\includegraphics[width=0.49\textwidth]{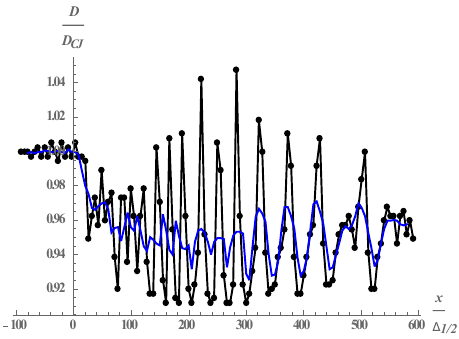}}  \label{speeds2}
	\hfill
	\subfloat[]{\includegraphics[width=0.55\textwidth]{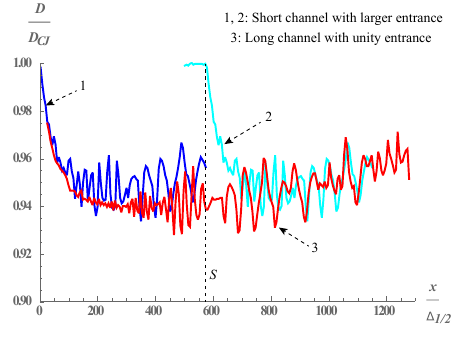}}  \label{speeds3}
	\caption{Evolution of a curved detonation in exponential channels with $K\Delta_{1/2} = 0.004$ \cite{Matei-Ramp}: (a) the superimposed detonation fronts at different instants for the long channel with unity entrance and the short channel with larger entrance height of 10$\Delta_{1/2}$; (b) the detonation speed measured along the straight bottom wall of the long channel and its local time-average; (c) the detonation speed measured along the straight bottom wall of the short channel and its local time-average and (d) comparison of local time-average speeds for the long channel and short one. Profile 2 is obtained by shifting profile 1 to the place $S$, where the long channel and short one have the same channel height. }
	\label{EntranceEffects}
\end{figure}

\textcolor{black}{In order to investigate whether the conclusion of short transient entrance effects can be generalized, we have further analyzed the numerical results of Radulescu and Borzou \cite{Matei-Ramp} for different entrance heights.  These new compiled results are shown in Fig.\ \ref{EntranceEffects}. In both simulations, a planar ZND detonation enters a diverging ramp with the constant  divergence rate, i.e., $\bar{K } = K \Delta_{1/2}=0.004$.  Two entrance heights are considered, 10$\Delta_{1/2}$ and 1$\Delta_{1/2}$, where $\Delta_{1/2}$ is the half reaction zone length. Further details can be found in  Ref.\ \cite{Matei-Ramp}.} 

\textcolor{black}{The evolution of the detonation front speed recorded along the straight bottom wall, as well as its running time average, are shown in Fig.\ \ref{EntranceEffects}b and Fig.\ \ref{EntranceEffects}c, respectively. From the running time average, it can be observed that it takes a length scale of approximately 200 (in the order of $1/0.004 = 250$) for detonations in both the long and short channels to achieve the quasi-steady state, despite their different entrance heights. In passing, one can also note that minor acceleration effects can be observed near the end of the channels, which is consistent with the slightly smaller curvature found near the end of ramps in Fig.\ \ref{KappaSteady} - an indication of the limitation of the quasi-1D assumption in the design of the exponential geometry previously discussed. }

\textcolor{black}{Moreover, results in Fig.\ \ref{EntranceEffects}d demonstrate that detonations in both the long and short channels follow the same evolution behaviors, since their running time average speed profiles collapse very well together. It can thus be concluded that the entrance heights have no significant influence on the evolution of detonations. The relaxation length scale appears to correlate well with the radius of curvature dictated by the specific geometry.}

\textcolor{black}{Note that our finding that the relaxation time and length both scale with $1/K$ is at odds with the quasi-steady state evolution of curved detonations described by the theory of Geometrical Shock Dynamics (GSD) \cite{bdzilDSD, lambournDSD}, which suggests that the relaxation time to steady state follows a process of diffusion of curvature on the surface of the front controlled by a time scale proportional to $h^2K$, where $h$ is the channel's thickness.  The quasi-steady GSD description requires that the detonation be much thinner than the observation length scales (comparable to the channel height $h$).  In our experiments, the detonation enters the curved channel when the channel height is comparable with the reaction zone thickness and comes to a steady state very rapidly.  This steady state is then maintained by the constant logarithmic derivative of the channel height.}

\subsection{The generalized ZND model}
\textcolor{black}{For the steady, inviscid, reacting quasi-1D flow behind the leading shock front of a detonation wave, the governing equations can be expressed in the frame of reference attached to the shock front as the following system of ordinary differential equations (ODEs) \cite{klein1995curved}}
\begin{subequations}
	\begin{align}
	\ode{}{x'}\left(\rho u A_{tot}\right) &=0 \label{continuity} \\
	\rho u \ode{u}{x'}+\ode{p}{x'} &=0 \label{momentum}\\
	\ode{}{x'}\left(h + \dfrac{1}{2}u^2 \right)&=0 \label{energy}\\
	u\ode{y_i}{x'} =\dfrac{W_i\dot{\omega}_i}{\rho}  & \quad (i=1,\cdots,N_s) \label{species}
	\end{align}
\end{subequations}
where $\rho, u, A_{tot}, p, h, y_i, W_i, \dot{\omega}_i,$ and $N_s$ are the mixture density, particle velocity, total cross-sectional area, pressure, enthalpy, species mass fraction, molecular weight, molar production rate of species $i$, and the total number of species. $x'$ is the distance from the shock front under the shock-attached reference. \textcolor{black}{The more convenient form of the above set of equations for computation is}
\begin{subequations}
	\begin{align}
	\ode{p}{t} &= -\rho u^2 \dfrac{\dot{\sigma}_{re}-\dot{\sigma}_A}{\eta} \label{dp/dt} \\
	\ode{\rho}{t} &=-\rho \dfrac{\dot{\sigma}_{re}-M^2\dot{\sigma}_A}{\eta} \\
	\ode{u}{t} &= u\dfrac{\dot{\sigma}_{re}-\dot{\sigma}_A}{\eta} \label{du/dt}\\
	\ode{y_i}{t} &=\dfrac{W_i\dot{\omega}_i}{\rho} \quad (i=1,\cdots,N_s)\\
	\ode{x'}{t}&= u \label{dx'/dt}
	\end{align} \label{simplified-eq}
	with
	\begin{align}
	\eta = 1-M^2, \qquad \dot{\sigma}_{re}=\sum\limits_{i=1}^{N_s}\left(\dfrac{W}{W_i}-\dfrac{h_i}{c_pT} \right)\ode{y_i}{t}, \quad \dot{\sigma}_A = \dfrac{u}{A_{tot}}\ode{A_{tot}}{x'}
	\end{align}
\end{subequations}
\begin{figure}[!htbp]
	\centering
	{\includegraphics[width=0.6\textwidth]{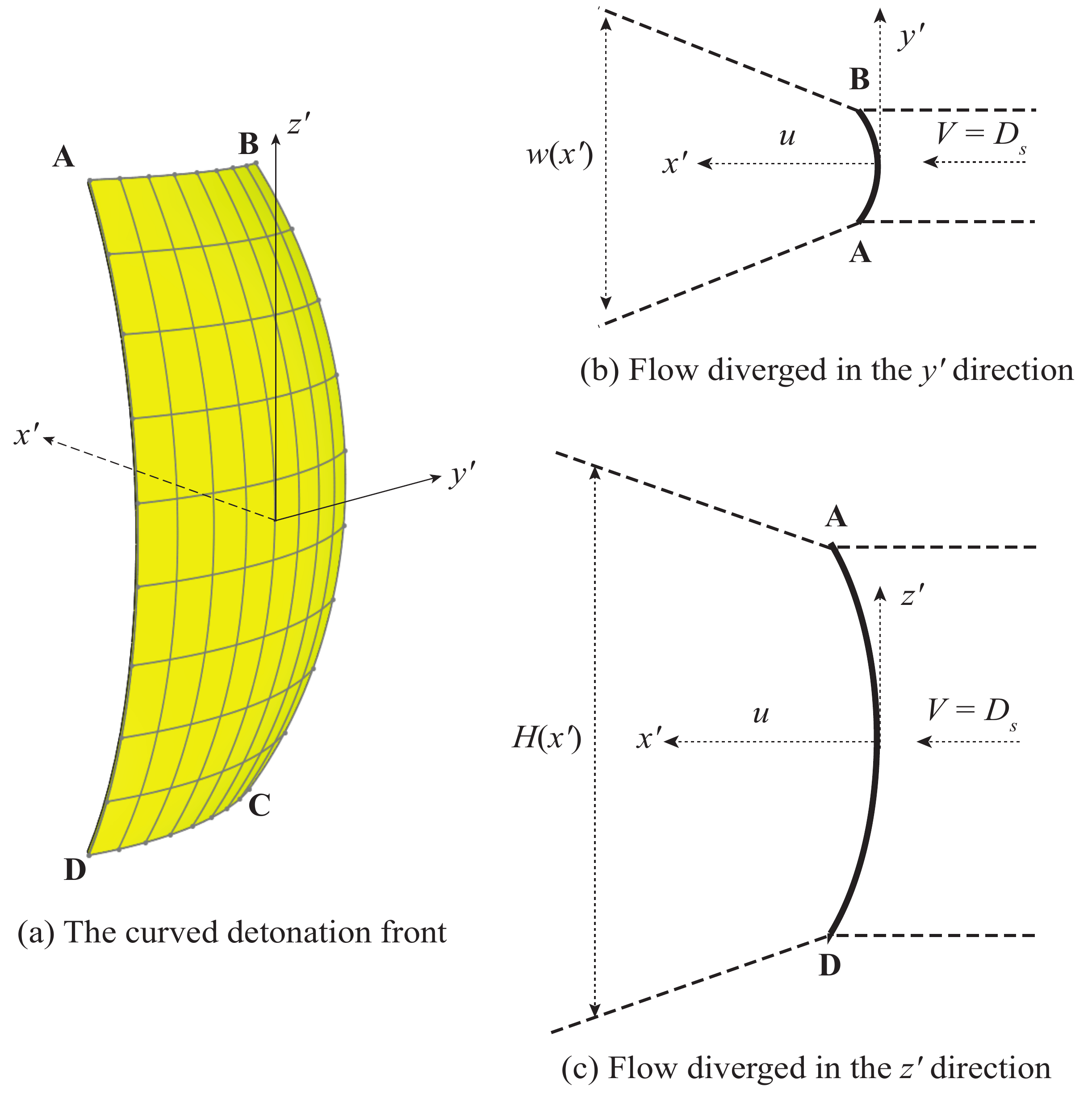}}
	\caption{\textcolor{black}{Sketch illustrating the flow diverged behind the curved detonation front under the shock attached reference. Note that the incoming flow propagates along the $x'$ direction.} } \label{ModelSketch}  
\end{figure}where $\dot{\sigma}_{re}$ is the thermicity of ideal gases, while $\dot{\sigma}_A$ is the rate of lateral strain. $M$, $h_i$, and $c_p$ are the local Mach number, specific enthalpy of species $i$, and mixture specific heat at constant pressure. For the detonation wave, the flow behind the leading shock starts from being subsonic and then accelerates towards the sonic condition due to the positive energy release of $\dot{\sigma}_{re}$. On the other hand, the strain rate of $\dot{\sigma}_A$ by the lateral flow divergence plays the opposite effect of decelerating this flow. As the flow may eventually become supersonic, a singular behavior would appear in Eqs.\ (\ref{dp/dt})-(\ref{du/dt}).  The local balance of these two competing effects, however, can effectively prevent this singular phenomenon by mathematically letting $\dot{\sigma}_{re}=\dot{\sigma}_A$ happen at the same point for $u=c$ ($M=1$). As a result, this so-called generalized CJ condition is satisfied with the steady reaction zone solution smoothly passing through the sonic point \cite{klein1995curved}. \textcolor{black}{ With this criterion, the simplified system of governing equations of Eq.\ (\ref{simplified-eq}) can be numerically solved.}

\textcolor{black}{The lateral strain rate experienced by the flow in the detonation attached frame can be related to two separate effects in our experiments: the divergence of the flow to boundary layers and that due to an enlarging channel dimension \cite{Matei-Ramp}. Consider a curved detonation front $ABCD$ in Fig.\ \ref{ModelSketch}a, where $x', y', z'$ are the space coordinates under the shock-attached system. The incoming flow (along the $x'$ direction) across the shock diverges in both $y'$ and $z'$ directions, as shown in Figs.\ \ref{ModelSketch}b and c, respectively. The lateral flow divergence behind the detonation can be related with the shock front curvature through the following expressions as \cite{fickett1979detonation, klein1995curved} 
\begin{subequations}
	\begin{align}
	\dfrac{1}{A_{tot}}\ode{A_{tot}}{x'} &= K_{eff}\left(\dfrac{D_s}{u}-1\right) \\ \dfrac{1}{w(x')}\ode{w(x')}{x'} &= K_{AB}\left(\dfrac{D_s}{u}-1\right) \\
	\dfrac{1}{H(x')}\ode{H(x')}{x'} &= K_{AD}\left(\dfrac{D_s}{u}-1\right)
	\end{align}
	where $D_s$ is the detonation speed. $K_{eff}$ is the effective curvature of the curved detonation front $ABCD$, while $K_{AB}$ and $K_{AD}$ are the curvature due to the diverging flow in directions of $y'$ and $z'$, respectively. Since the total cross-sectional area $A_{tot}$ is $A_{tot} = H(x')\times w(x')$, we can thus have 
	\begin{align}
	\dfrac{1}{A_{tot}}\ode{A_{tot}}{x'} = \ode{}{x'}\left(lnA_{tot}\right) = \dfrac{1}{H(x')}\ode{H(x')}{x'} + \dfrac{1}{w(x')}\ode{w(x')}{x'}
	\end{align}
	Therefore, we can further obtain
	\begin{align}
	K_{eff} = K_{AD} + K_{AB}  
	\end{align}
\end{subequations}
\label{Keff}which indicates that for the curved detonation front $ABCD$ in Fig.\ \ref{ModelSketch}a, its total curvature of $K_{eff}$ includes two parts, one due to the flow diverged in the $y'$ direction while the other one comes from the  $z'$ direction. Similarly, for the curved detonation in our exponential channels, its total flow divergence also includes two parts, the one from the geometrical divergence of the exponentially diverging channel in the height direction and the other due to divergence of the flow rendered by the boundary layer growth on the channel side walls. The effective curvature $K_{eff}$, experienced by detonations in the exponential geometry, can thus be expressed as
\begin{align}
K_{eff} = K + \phi_{BL}
\end{align} 
where $K$ is the logarithmic area divergence rate of the geometry. For the large ramp, $K=2.17$ m$^{-1}$, while for the small one, $K=4.34$ m$^{-1}$. $\phi_{BL}$ represents the contribution of the boundary-layer-induced losses from the channel width direction.
}
\subsection{The experimental $D(\kappa)$ curves}

\begin{figure}[!htbp]
	\centering
	{\includegraphics[width=0.85\textwidth]{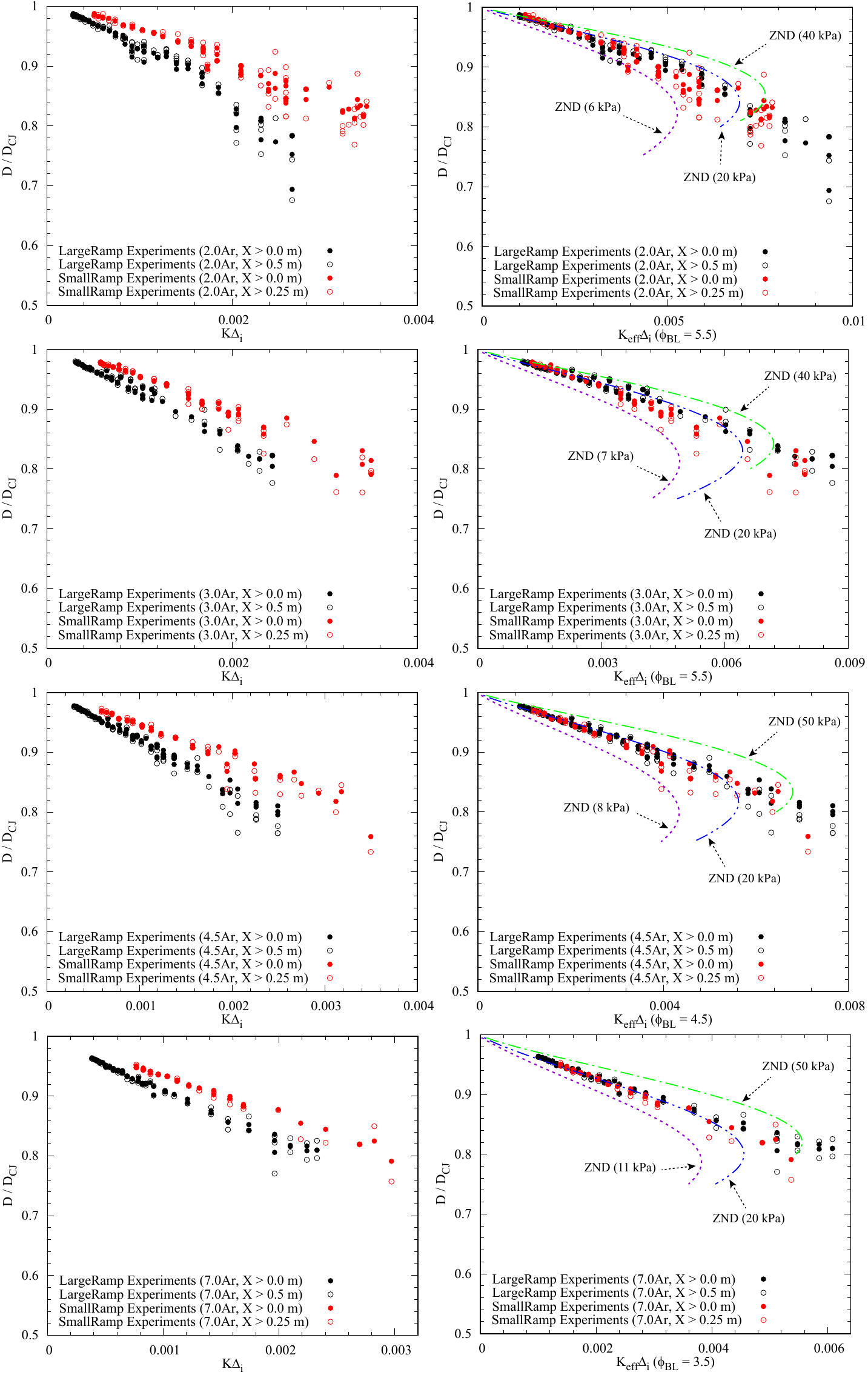}}
	\caption{\textcolor{black}{The non-dimensional $D-K$ characteristic relationships obtained experimentally and predicted from the generalized ZND model using the detailed San Diego chemical mechanism.} } \label{H2-O2-Ar-DKandKeff}  
\end{figure}

Instead of modelling \textcolor{black}{ the boundary-layer-induced lateral flow divergence}, Radulescu and Borzou \cite{Matei-Ramp} directly evaluated this loss rate \textcolor{black}{of  $\phi_{BL}$} from experiments by analytically comparing the experimental data of two ramps with two underlying assumptions: (1) detonations propagating inside the exponentially diverging channel of different expansion ratios have the same constant $\phi_{BL}$ since the channel's dimension  of the width is unchanged; (2) for the same mixture, it has a unique relation between the velocity deficits and the losses. As a result, the effective \textcolor{black}{curvature $K_{eff}$ of the global front } can be calibrated by collapsing together the experimental $D(\kappa)$ curves of detonations in the large and small ramp experiments, and then the loss rate $\phi_{BL}$ due to boundary layers can be derived \cite{Matei-Ramp}. Figure\ \ref{H2-O2-Ar-DKandKeff} shows the experimentally obtained  $D(\kappa)$ curves, characterizing the relationships between the detonation velocity (normalized by the ideal CJ speed) and the lateral flow divergence, for all the mixtures involved in this study. Note that the abscissa is the non-dimensional loss obtained by multiplying \textcolor{black}{the curvature} with the ZND induction zone length $\Delta_i$, which is taken as the distance between the CJ detonation front and the location of its peak thermicity. It was calculated by using the Shock and Detonation Toolbox (SDToolbox) \cite{kao2008numerical} under Cantera's framework \cite{cantera} with the San Diego chemical reaction mechanism (Williams) \cite{sandiego2014}. This normalization method is the same as that adopted by Radulescu and Borzou \cite{Matei-Ramp}. \textcolor{black}{Moreover, theoretical works have demonstrated that, for a specific mixture, there exists a unique relationship between the curved detonation velocity deficit and the curvature normalized by the induction zone length  \cite{klein1995curved, he1994direct, klein1993relation}.} In  Fig.\ \ref{H2-O2-Ar-DKandKeff}, the graph in the right column is the collapsed $D/D_{CJ}-K_{eff}\Delta_i$ correlation, after calibrating the effective curvature $K_{eff}$ from the $D/D_{CJ}-K\Delta_i$ curve in the corresponding left graph, by including the boundary layer effects.  In the work of Radulescu and Borzou \cite{Matei-Ramp}, the same unique value of $\phi_{BL}=5.5$ m$^{-1}$ was found to permit collapsing all data of two ramps in both mixtures of 2C${_2}$H${_2}$/5O$_2$/21Ar and C${_3}$H${_8}$/5O$_2$. However, the present study found that $\phi_{BL}$ can be varied for different mixtures, with  5.5 m$^{-1}$ for 2H${_2}$/O$_2$/2.0Ar and 2H${_2}$/O$_2$/3.0Ar, while 4.5 m$^{-1}$ and 3.5 m$^{-1}$ for 2H${_2}$/O$_2$/4.5Ar and 2H${_2}$/O$_2$/7.0Ar, respectively.

\textcolor{black}{In the recent works of Kudo et al. \cite{kudo2011oblique} and Nakayama et al. \cite{nakayama2013front, nakayama2012stable} of curved gaseous detonations in rectangular-cross-section curved channels, they also obtained the meaningful characteristic $D(k)$ curves, which are the first extension of the $D(\kappa)$ theory experimentally from condensed-phase detonations to gas-phase detonations. When the detonation cell size was small enough, i.e., in the order of 0.1$\sim$1.0 mm, they were able to obtain the curved detonations in quasi-steady conditions. An interesting question that arises is why the boundary layer effects can be negligible in their works while appear to be significant in the present work, since their channel depth of 16 mm is in the same order with that of the current exponential channels. This can be first clarified by the fact that the cell sizes of detonations in the present study are comparable to, if not larger than, the channel width, while in works of Nakayama et al., the cell size is much smaller. As a result, detonations of much longer characteristic reaction zones in the present experiments  experienced more significant boundary-layer-induced losses. The other explanation can be interpreted with the detonation loss induced by the geometry. The exponential-geometry-induced mean front curvature in the present work has been found to be comparable to that due to boundary layers. However, in works of Nakayama et al., the local curvature induced by the geometry ranges from 20 $\sim$ 200 m$^{-1}$ (evaluated from Fig.\ 7 of Ref.\ \cite{nakayama2013front}), which is much larger than the boundary-layer-induced loss rate estimated from current experiments. Therefore, as a result of the difference between geometries and initial conditions, the boundary layer can play the role of varied significance.  }

\begin{figure}[!htb]
	\centering
	{\includegraphics[width=0.6\textwidth]{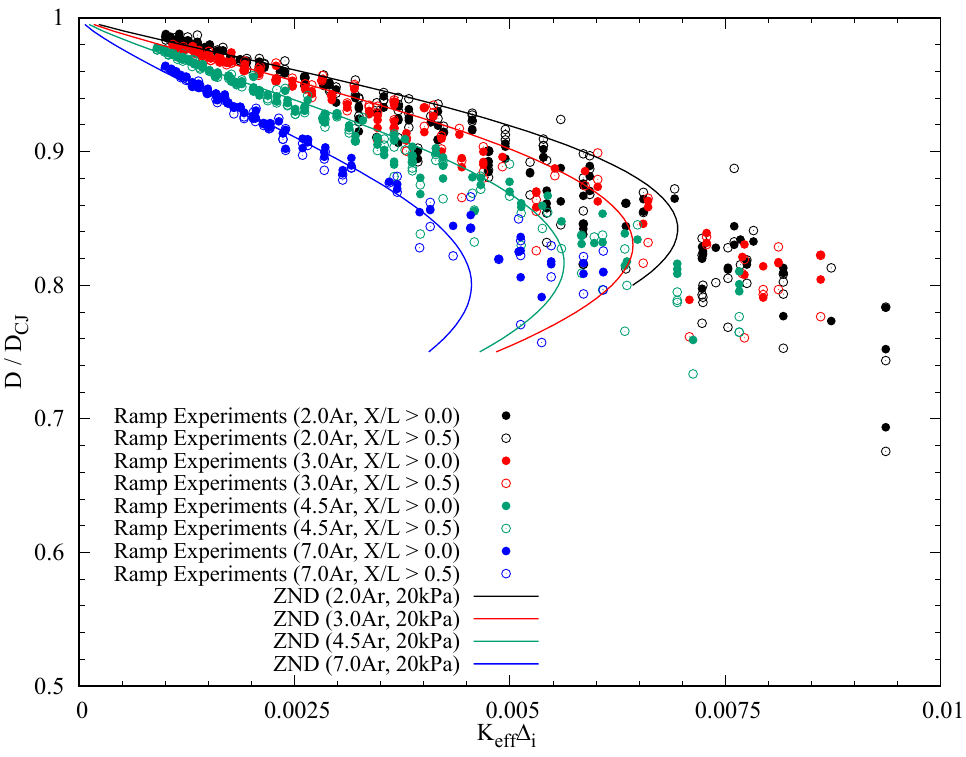}}
	\caption{\textcolor{black}{The $D/D_{CJ}-K_{eff}\Delta_i$ characteristic curves obtained from experiments and predicted from the quasi-1D ZND model for different mixtures.} } \label{DKeff-Combined}  
\end{figure}

\begin{figure}[!htb]
	\centering
	{\includegraphics[width=0.5\textwidth]{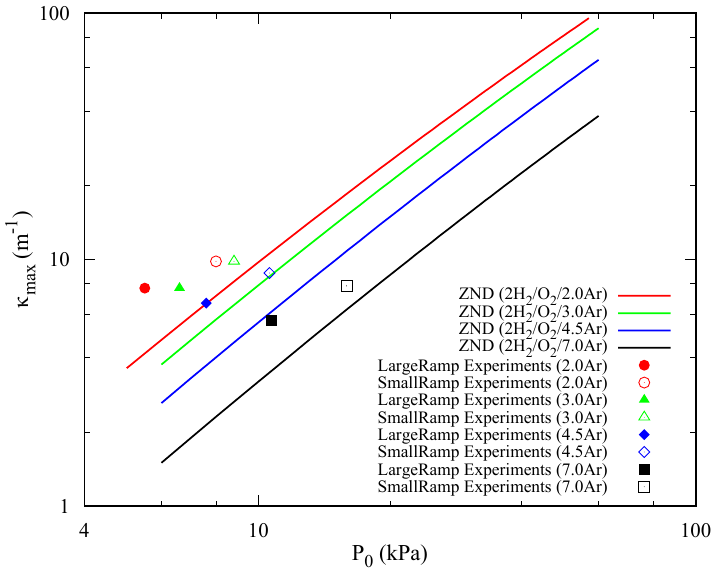}}
	\caption{\textcolor{black}{Critical curvature as a function of initial pressures obtained from experiments and predicted from the quasi-1D ZND model for different mixtures.}} \label{H2-O2-Ar-CriticalCurvature}  
\end{figure}

\subsection{Comparisons with the generalized ZND model}
\subsubsection{The ZND model predicted $D/D_{CJ}-K_{eff}\Delta_i$ curves}
To start with, the theoretically predicted relationship between the \textcolor{black}{effective curvature} $K_{eff}$ and the detonation speed has been obtained \textcolor{black}{by solving the ODE system of Eq.\ (\ref{simplified-eq}), with the developed custom Python code \cite{Matei-Ramp} working under the framework of SDToolbox and Cantera. The San Diego reaction mechanism was applied for describing the realistic chemical kinetics.} These calculations were conducted at \textcolor{black}{three initial pressures covering the experimental range}. Normalizing the varied divergence rates $K_{eff}$ by the ZND induction zone length $\Delta_i$ at these pressures permits getting the theoretical $D/D_{CJ}-K_{eff}\Delta_i$ curves, \textcolor{black}{which have been shown in Fig.\ \ref{H2-O2-Ar-DKandKeff}}.  \textcolor{black}{These comparisons show that the experiments are generally in very good agreement with the extended ZND model predictions for small and moderate lateral strain, except near the limit.} Detonations from experiments were able to propagate beyond the maximum lateral strain predicted by the steady ZND model, with larger velocity deficits, thus demonstrating the role of cellular structures in enhancing the detonability of gaseous detonations  \cite{radulescu2018detonation}. \textcolor{black}{The $D/D_{CJ}-K_{eff}\Delta_i$ characteristic relationships of both the experiments and ZND model predictions for all the mixtures, as summarized in Fig.\ \ref{DKeff-Combined}, demonstrate that detonations with less argon dilutions can propagate with more losses.}  It is thus indicative of the effects of argon dilutions in reducing the detonability of  H${_2}$/O${_2}$/Ar detonations, consistent with the computations by Klein et al. \cite{klein1995curved}.

\subsubsection{\textcolor{black}{The ZND model predicted critical curvature $\kappa_{max}$}}
\textcolor{black}{The effects of argon dilutions in reducing the detonation detonability are more evident in Fig.\ \ref{H2-O2-Ar-CriticalCurvature}, which illustrates the relationship between the initial pressures and the ZND model predicted critical curvature $\kappa_{max}$, above which detonations are not possible at a certain initial pressure. With the reduction of the mixture sensitivity by lowering the initial pressure, the critical curvature also decreases with the reduced detonability. Moreover, the results in Fig.\ \ref{H2-O2-Ar-CriticalCurvature} further demonstrate that detonations in experiments can propagate with a larger critical curvature than that predicted by the generalized ZND model, suggesting the higher detonability of real detonations in experiments.}

\subsubsection{The ZND model predicted detonation speed as a function of initial pressures}

\begin{figure}[!htbp]
	\centering
	{\includegraphics[width=0.85\textwidth]{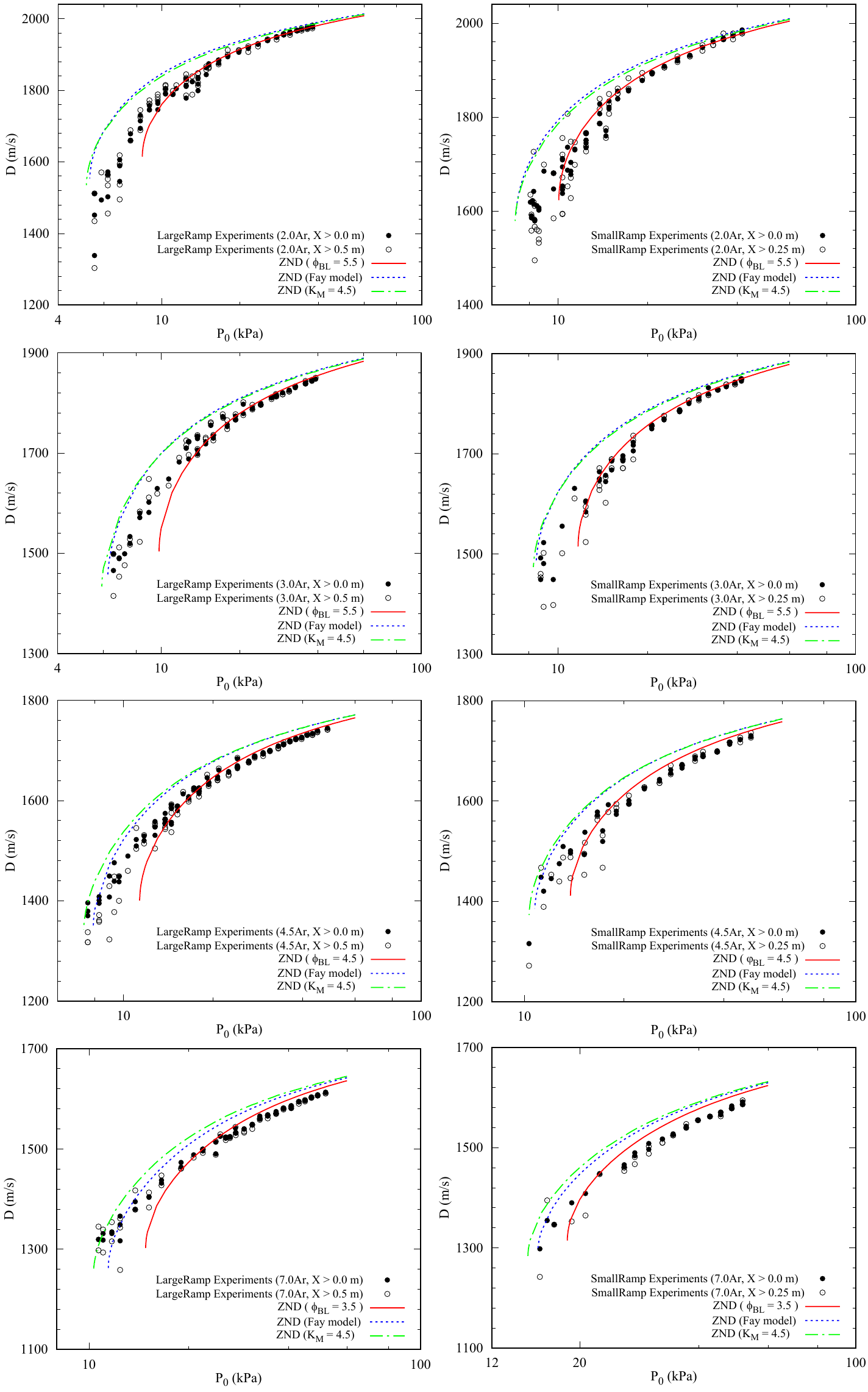}}
	\caption{\textcolor{black}{Comparisons of the experimentally obtained average speeds with the generalized ZND model predictions for various mixtures; the red line adopts the experimentally obtained constant divergence rate of $\phi_{BL}$ due to the boundary layer for the ZND model prediction, while the broken blue line and green line are the predictions made with the turbulent and laminar boundary layer assumptions, respectively.}} \label{H2-O2-Ar-ZND}  
\end{figure}

\textcolor{black}{Finally, the theoretically predicted detonation speed as a function of initial pressures has also been calculated, as shown in Fig.\ \ref{H2-O2-Ar-ZND} illustrating the comparisons of the experiments with the predictions made with the real chemistry. Such ZND model predicted relationships were obtained through three approaches for evaluating the boundary-layer-induced flow divergence, either directly from experiment, or by modeling the negative displacement thickness in the case of turbulent or laminar boundary layers. \\
(1) The first method assumed the constant boundary-layer-induced loss rate $\phi_{BL}$, which has been directly calibrated from experiments. The lateral strain rate $\dot{\sigma}_A$ can thus be expressed as
\begin{align}
\dot{\sigma}_A &= \left(D_s - u\right)\left(K+\phi_{BL}\right)
\end{align} 
(2) According to Fay's theory on boundary layer mechanism, its effect can be modelled by inviscid flow in a streamtube with a negative displacement of the boundary layer, as a result of the boundary layer acting as a mass sink by removing the mass from the core flow \cite{fay1959two}. For the present experimental configuration, since the channel height is much larger than the channel width, the boundary layer effects on the top and bottom curved wall can be reasonably neglected. The effective width of the diverged flow behind the detonation is thus given by $w(x') = w + 2\delta^*(x')$, where $\delta^*(x')$ is the boundary layer negative displacement thickness . As a result, the rate of flow divergence due to the boundary layer growth on side walls can be evaluated as 
\begin{align}
\dfrac{1}{w(x')}\ode{w(x')}{x'}=\dfrac{2}{\left(w+2\delta^* \right)}\ode{\delta^*}{x'} \label{area-divergence-eq}
\end{align}}In Fay's work, he adopted Gooderum's empirical turbulent boundary layer thickness relation \cite{gooderum1958experimental} as the displacement thickness behind the detonation wave. The relation is given by 
\begin{align}
\delta^*(x') = 0.22\left( x' \right) ^{0.8}\left(\dfrac{\mu_e }{\rho_0 D_s} \right)^{0.2} \label{Fay}
\end{align}        
where $\mu_e$ and $\rho_0$ are the post-shock state viscosity and the initial density, respectively. This turbulent boundary layer displacement thickness relation has then been applied in a large number of subsequent works investigating detonation velocity deficits \cite{dove1974velocity, murray1985influence,chao2009detonability, camargo2010propagation,ishii2011detonation,zhang2015detonation,gao2016experimental,ZHANG2017193}. The boundary-layer-induced flow divergence rate is thus a function of the distance behind the leading shock. Consequently, using this turbulent boundary layer displacement thickness relation, the lateral strain rate $\dot{\sigma}_A$ is
\begin{align}
\dot{\sigma}_A &= \left(D_s - u\right)K + u\left(\dfrac{0.352}{\text{w}+2\delta^*}\right)\left(\dfrac{\mu_e }{\rho_0 D_s} \right)^{0.2} \left(\dfrac{1 }{x'} \right)^{0.2}
\end{align}\textcolor{black}{(3) For laminar boundary layers, the displacement thickness was obtained by solving Mirels' compressible laminar boundary layer equations \cite{mirels1956boundary} (see the Appendix)
\begin{align}
\delta^*(x') =  4.5\sqrt{\dfrac{\mu_e x'}{\rho_eu_e}} 
\end{align}
where we have $\rho_eu_e = \rho_0 D_s$ via the mass conservation. The lateral strain rate $\dot{\sigma}_A$ then is 
\begin{align}
\dot{\sigma}_A &= \left(D_s - u\right)K + u\left(\dfrac{4.5}{\text{w}+2\delta^*}\right)\left(\dfrac{\mu_e }{\rho_0 D_s} \right)^{0.5} \left(\dfrac{1 }{x'} \right)^{0.5}
\end{align}	}

\begin{figure}[]
	\centering
	{\includegraphics[width=0.5\textwidth]{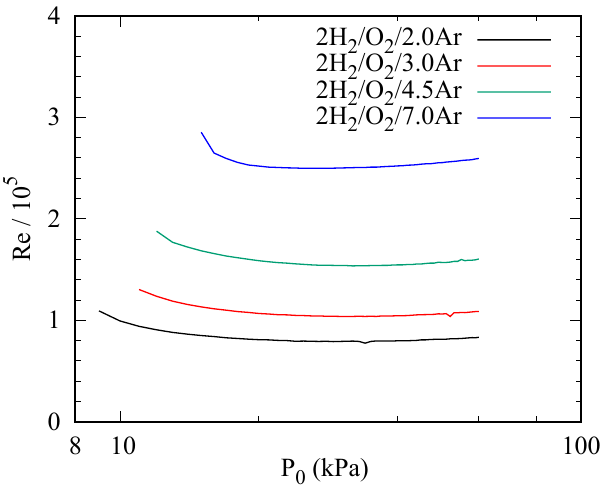}}
	\caption{Reynolds number $Re = \left(\rho_s u_s/\mu_s\right)x_H$ for H${_2}$/O$_2$/Ar detonations with the same lateral flow divergence rate as those in large ramp experiments under varied initial pressures.   } \label{ReynoldsNumber}  
\end{figure}

\textcolor{black}{The comparisons in Fig.\ \ref{H2-O2-Ar-ZND} show that the ZND model predictions, obtained with Fay's turbulent boundary layer displacement thickness relation and Mirels' laminar boundary layer solutions, both have the very close results. They can relatively well predict the experiments. And surprisingly, they also appear to capture the near-limit detonation dynamics quite well, including both the limit pressures and velocity deficits.}   Moreover, we also estimated the Reynolds number $Re = \left(\rho_s u_s/\mu_s\right)x_H$, as shown in Fig.\ \ref{ReynoldsNumber}, for  H${_2}$/O$_2$/Ar detonations in large ramp experiments. Note that $\rho_s$, $u_s$, and $\mu_s$ are the post-shock parameters, i.e., density, particle velocity (in the lab frame reference), and viscosity, while $x_H$ is taken as the characteristic hydrodynamic thickness between the leading shock and the CJ sonic surface. The results from Fig.\ \ref{ReynoldsNumber} indicate that the boundary layer behind H${_2}$/O$_2$/Ar detonations is probably laminar, since the Reynolds number is smaller than the critical one of \textcolor{black}{$Re_c \approx 0.5\times 10^6\sim 4.0 \times 10^6$ for the shock-induced boundary-layer transition from laminar to be turbulent \cite{gooderum1958experimental, ra1960boundary, mirels1964shock, petersen2003improved}}. This finding concurs with the previous conclusion of Liu and Glass \cite{liu1983laminar} and Damazo et al. \cite{damazo2012boundary} that the boundary layer behind stoichiometric hydrogen/oxygen detonations is laminar. \textcolor{black}{Therefore, Fay's turbulent boundary layer displacement thickness relation for evaluating the boundary-layer-induced flow divergence is questionable. It is not clear at present why this turbulent boundary layer assumption somehow can work well for predicting the experiments.}

On the other hand, \textcolor{black}{it can also be clearly seen from Fig.\ \ref{H2-O2-Ar-ZND} that} the experiments are in excellent agreement with the predictions, made with the constant boundary-layer-induced loss rate of $\phi_{BL}$ directly evaluated from experiments, for detonations with small and moderate velocity deficits well above the limit. While approaching the limit, these predictions deteriorate. Real detonations can propagate at lower initial pressures, where the steady ZND model predicts failure. The first reason for this discrepancy  near the limit can be attributed to the possible promotion mechanism of strong reactive transverse waves, especially the transverse detonations, which help extend the propagation limits to lower pressures than that predicted by the generalized ZND model neglecting cellular structures. Secondly, the assumption of a constant global curvature for the leading shock front, well above the limit, is not applicable to detonations near the limit, as can be concluded from observations of the structures of detonation fronts at relatively low pressures, e.g., see Fig.\ \ref{superposition2}. It thus results in failure of the theoretical predictions for \textcolor{black}{these near-limit detonations from experiments}.

\begin{figure}[]
	\centering
	{\includegraphics[width=0.7\textwidth]{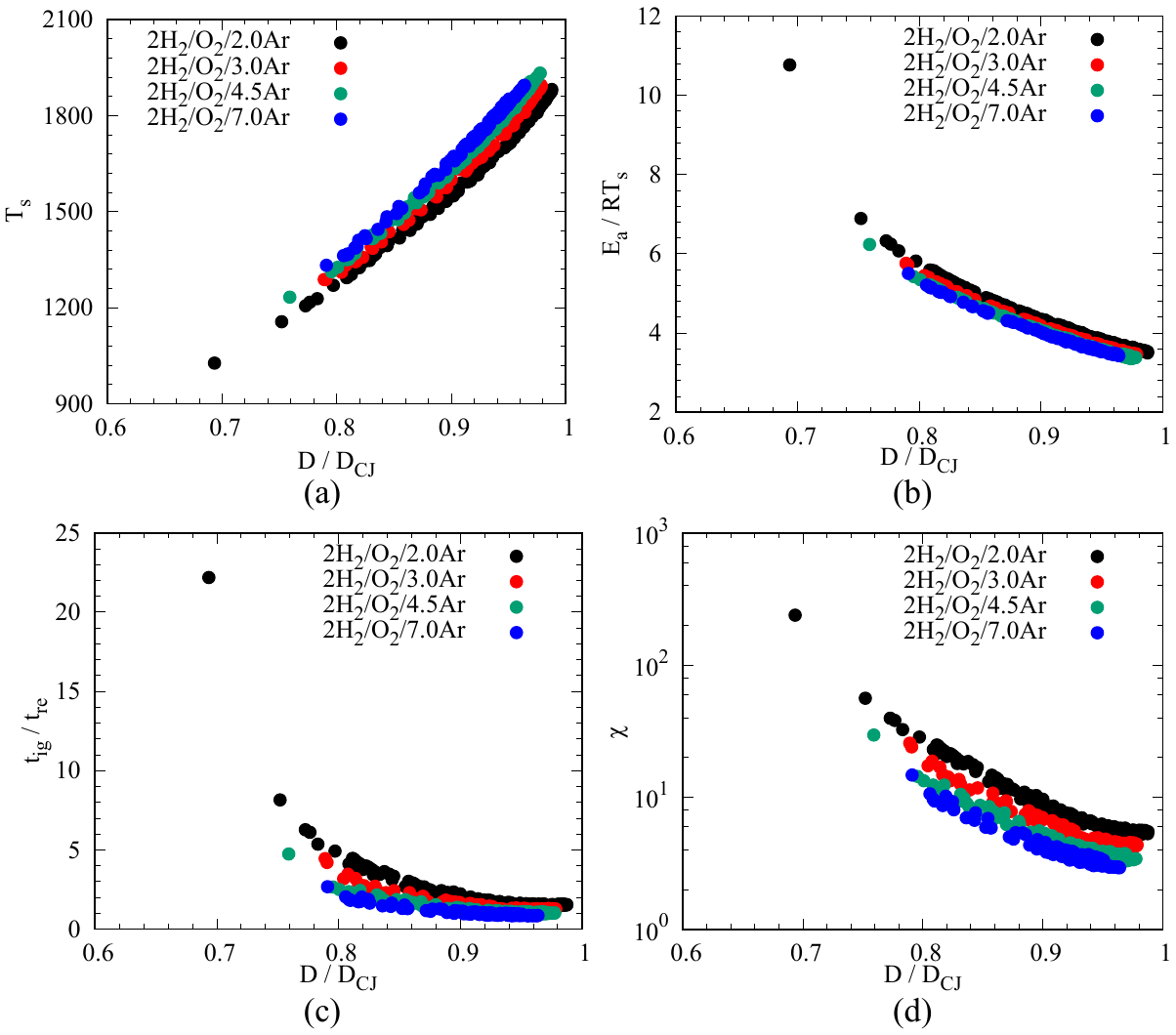}}
	\caption{Parameters of (a) post-shock temperature $T_s$, (b) reduced activation energy $\theta = E_a/RT_s$, (c) the ratio of ignition time to reaction time, and (d) the stability parameter $\chi$ as a function of the shock speed normalized with the CJ velocity. The ignition time here is defined by the time to peak thermicity,  while the reaction time is defined as the inverse of the maximum thermicity, as proposed by Radulescu \cite{radulescu2003propagation}. They were calculated with the constant volume explosion at the mean propagation speed with the real chemistry.} \label{parameters}  
\end{figure}

\textcolor{black}{The observed differences between the ZND model prediction and experiments observed near the limits may also be affected by the enhanced instability of attenuated detonations with lower shock temperatures and longer ignition delays compared to the reaction time scales}. Radulescu \cite{radulescu2003propagation} introduced the parameter $\chi$ for characterizing the stability of detonations under different thermodynamic conditions. Detonations in mixtures of higher $\chi$ are more unstable to perturbations in the reaction zones than those with lower $\chi$. The mathematical expression of  $\chi$ is 
\begin{equation}
\chi = \left(\dfrac{E_a}{RT_s}\right)\left(\dfrac{t_{ig}}{t_{re}} \right)
\end{equation} 
where $E_a/RT_s$ is the usual non-dimensional activation energy, $T_s$ is the temperature behind the leading shock, $R$ is the specific gas constant, and $t_{ig}/t_{re}$ is the ratio of ignition to reaction time. Figure\ \ref{parameters} shows, for all the mixtures, the relationships of these parameters as a function of the leading shock velocity, normalized with the ideal CJ speed. Of noteworthy is that each solid circle represents one experiment, and the shock speed corresponds to the experimentally measured mean propagation velocity \textcolor{black}{over the whole ramps}. One can observe that decreasing the detonation front speed by increasing the velocity deficits results in the reduced post-shock temperature, increased activation energy, and larger ratios of the ignition time relative to the reaction time, thereby enhancing the sensitivity of the reaction rates to relaxations in the reaction zones. As such, detonations tend to become more unstable with increased $\chi$. The considerable increment of $\chi$ , however, occurs when the velocity deficit is relatively large, especially near the limit, as can be observed from Fig.\ \ref{parameters}d. It can thus be speculated that the significantly increased instability, as a result of considerably large velocity deficits, leads to the incapability of the extended ZND model to predict detonation dynamics at relatively low initial pressures near the limit.  

\subsection{Why can the steady 1D ZND model predict the H${_2}$/O$_2$/Ar cellular detonation dynamics?}

The above analysis has quantitatively showed that the generalized ZND model with lateral strain rate can predict very well the experiments of H${_2}$/O${_2}$/Ar cellular detonations, except \textcolor{black}{some departures} for the near-limit detonation dynamics. \textcolor{black}{The question that arises is what results in such excellent predictability, given the detonation structure is always cellular.}

\begin{figure}[]
	\centering
	{\includegraphics[width=0.5\textwidth]{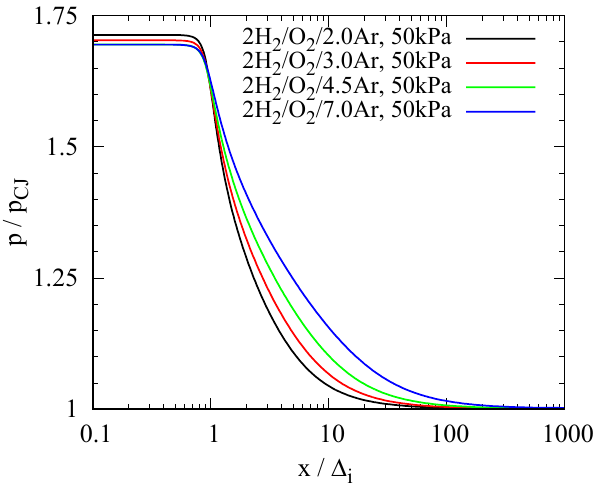}}
	\caption{\textcolor{black}{The ZND structure of CJ detonations: pressure profile as a function of the distance normalized by the induction zone length.} } \label{H2-O2-Ar-ZND-Structure-CJ}  
\end{figure}

\textcolor{black}{A tentative answer comes from the analysis of the relevant length and time scales in the detonation cellular structure.  For hydrogen-oxygen-argon detonations, the induction time scales and length scales are much shorter than the global reaction zone, as can be seen from the ZND calculations of Fig.\ \ref{H2-O2-Ar-ZND-Structure-CJ}.  90$\%$ of energy release occurs at a distance an order of magnitude longer than the induction zone thickness. Since the reaction zone is not thermally sensitive, spatial variations in the induction times associated with the cellular structure do not appreciably change the reaction zone structure.  Since the dynamics are controlled by the acoustic time over the whole reaction zone structure, these are expected to be insensitive to the induction zone variations, which are much faster.  For hydrocarbon detonations studied by Radulescu and Borzou \cite{Matei-Ramp}, the opposite is true and the reaction zone thicknesses are much shorter as compared to the induction zone thickness.  For these mixtures, the ZND model fails to capture the dynamics.}


\section{Conclusion}
\label{Conclusion}
In the present study, experiments of very regular cellular detonations, propagating inside the exponentially diverging channels with two different constant area divergence rates, were performed in mixtures of stoichiometric H${_2}$/O$_2$ of varied argon dilutions. The propagation characteristics were demonstrated and analyzed in detail. The results showed that detonations, well above the limit, were uniformly curved with small-sized cellular structures of constant cell sizes, and can be reasonably assumed to propagate in quasi-steady state at the macro-scale with a constant \textcolor{black}{mean front curvature}. The experimentally measured cell sizes showed a strong dependence on the detonation velocity deficits. When the mean propagation speed dropped to about $0.8 D_{CJ}$, detonations started to propagate in the mechanism of single-headed detonations, organized with a distinctive transverse detonation, with the detonation cell size larger than the ideal one in the order of $10 \sim 15$. The stabilization mechanism of single-headed detonations with saliently enlarging cells suggests a higher growth rate of the curved detonation front area than that of the intrinsic transverse instability. The Go/No-Go phenomenon has also been demonstrated for detonations near the very limit, due to its stochastic property. Also, both the reactive and non-reactive transverse waves were observed for detonations propagating at low pressures near the limit. 

The experimental $D/D_{CJ}-K_{eff}\Delta_i$ curves were obtained by collapsing the $D/D_{CJ}-K\Delta_i$ relationships of both the large ramp and small ramp. As a result, the equivalent \textcolor{black}{loss rate} $\phi_{BL}$ due to boundary layers were directly derived. It was found that detonations of lower argon dilutions can propagate with more losses, indicating the effects of argon addition in decreasing the detonability of H${_2}$/O$_2$/Ar detonations. \textcolor{black}{The same conclusion can also be drawn from the decrease of critical curvature with the increasing argon dilutions. The variation of velocity deficits with the flow divergence was found in very good agreement with the predictions made with steady ZND model with lateral strain rate, for small and moderate divergence and velocity deficits. Nevertheless, the model under-predicted the limiting lateral strain and velocity deficits for detonations near the limit, thus illustrating the effects of cellular structures in enhancing the detonability. } 

Furthermore, comparisons between the experimentally obtained average speeds and the generalized ZND model predictions were performed in terms of different initial pressures. \textcolor{black}{The results showed that predictions made with the experimentally obtained constant loss rate $\phi_{BL}$ can excellently predict the experiments, except some departures for the near-limit detonations. On the other hand, the predictions made with Fay's turbulent boundary layer displacement thickness relation and Mirels' laminar boundary layer solutions both can relatively well predict the experiments. However, the estimated Reynolds number casts doubt on the turbulent boundary layer assumption, while appears to support the finding that the boundary layer behind hydrogen/oxygen detonations is laminar.} Finally, the promotion effects by the strong reactive transverse waves or transverse detonations, \textcolor{black}{the limitation of the quasi-1D assumption  near the limit, and the considerably increased instability were proposed for clarifying the discrepancies between the theoretical predictions and experiments for the near-limit detonations. The much shorter induction zone length and time scales as compared to the global reaction zone were proposed for clarifying the predictability of the real H${_2}$/O$_2$/Ar cellular detonation dynamics by the extended 1D ZND model with lateral strain rate.}

\section*{Acknowledgments}
\label{Acknowledgments}
The authors acknowledge the financial support from the Natural Sciences and Engineering Research Council of Canada (NSERC) through the Discovery Grant "Predictability of detonation wave dynamics in gases: experiment and model development". The authors would also like to thank Jiaxin Chang, Maxime La Fleche, and Yongjia Wang for help in conducting the experiments, and thank S.M. Lau-Chapdelaine for the code for post-processing the shadowgraph and schlieren photographs.

\section*{Appendix: Mirels' shock-induced compressible laminar boundary layer solutions}
\begin{figure}[]
	\centering
	{\includegraphics[width=0.5\textwidth]{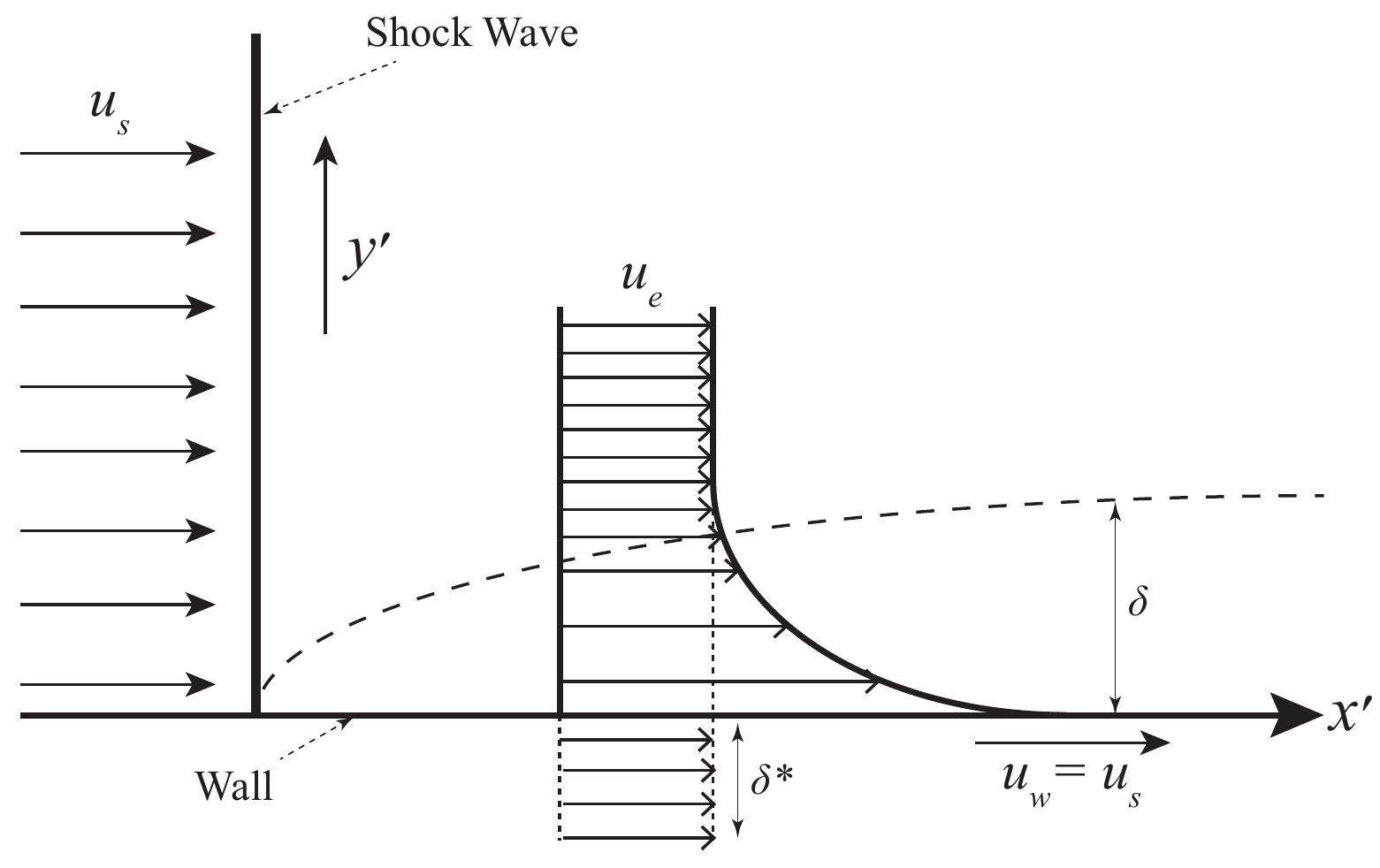}}
	\caption{Sketch of the boundary layer velocity profile behind the shock in the shock frame of reference. $u_s$ is the shock speed in the laboratory frame, while $u_e$ is the post-shock flow velocity outside the boundary layer in the shock frame. $\delta$ and $\delta^*$ represent the boundary layer thickness and displacement thickness, respectively.  } \label{Sketch-BL}  
\end{figure}

Here we present the solutions for the shock-induced compressible laminar boundary layer, which  has been investigated in detail in the 1950s by Mirels \cite{mirels1956boundary}. By assuming a laminar flow with zero pressure gradient ($dp/dx' = 0$), the boundary-layer equations can be expressed under the reference of the shock as

\begin{subequations}
	\begin{align}
	\pde{\left(\rho u\right)}{x'} + \pde{\left(\rho \upsilon\right)}{y'} &= 0 \\
	u\pde{u}{x'} + \upsilon\pde{u}{y'} & = \dfrac{1}{\rho} \pde{}{y'}\left(\mu \pde{u}{y'}\right)\\
	\rho c_p \left(u\pde{T}{x'}+ \upsilon\pde{T}{y'}\right) &= \pde{}{y'}\left(k\pde{T}{y'}\right)+ \mu \left(\pde{u}{y'}\right)^2 \\
	p &= \rho R T 
	\end{align} \label{BL-eq}with the corresponding boundary conditions behind the shock ($x'>0$) as
	\begin{align}
	u(x', 0) &= u_w \qquad u(x', \infty) = u_e \notag \\
	\upsilon(x', 0) &= 0 \\ 
	T(x', 0) &= T_w \qquad T(x', \infty) = T_e \notag 
	\end{align}
\end{subequations}
where $\left(x',y'\right)$ are the space coordinates in the shock frame of reference, whose configuration is shown in Fig.\ \ref{Sketch-BL}. The gas thermal conductivity $k$ and dynamic viscosity $\mu$ are assumed to scale linearly through the boundary layer with temperature. Note that the wall temperature $T_w$ is assumed constant, i.e., 298 K in the present study.  $u_w$ is the wall velocity with respect to the shock, $u_e$ and $T_e$ are the external flow velocity and temperature outside the boundary layer, as illustrated in Fig.\ \ref{Sketch-BL}. One more assumption involves the specific heat capacity $c_p$ and the Prandtl number $Pr$, which are also considered to be constant and equal $c_{p, w}$ and $Pr_w$.

By automatically satisfying the mass conservation in the steady two-dimensional boundary layer system, a transformation is made such that  
\begin{subequations}
	\begin{align}
	\dfrac{\rho u}{\rho_w} = \pde{\psi}{y'} \qquad \dfrac{\rho \upsilon}{\rho_w} = - \pde{\psi}{x'}
	\end{align}with the stream function $\psi$ and similarity parameter $\eta$ defined by
	\begin{align}
	\psi &= \sqrt{2u_ex' \nu_w}f(\eta) \\
	\eta &= \sqrt{\dfrac{u_e}{2x'\nu_w}}\intl{0}{y'} \dfrac{T_w}{T(x',y')}dy'
	\end{align}
\end{subequations}
where $\rho_w$ and $\nu_w$ are the density and kinetic viscosity at the wall, respectively. As such, the momentum equation can be finally transformed to
\begin{subequations}
	\begin{align}
	f''' + ff'' &= 0  \notag \\
	f(0)=0 \qquad f'(0) &= u_w/u_e \qquad f'(\infty) = 1 \label{ode-f} 
	\end{align}with the energy equation reduced to 
	\begin{align}
	\left(\dfrac{T}{T_e}\right)'' + Pr  \left(\dfrac{T}{T_e}\right)' f&= -Pr\dfrac{ {u^2_e}}{c_{p}T_e}\left(f''\right)^2 \notag \\
	\dfrac{T(x',0)}{T_e} = \dfrac{T_w}{T_e} &\qquad \dfrac{T(x', \infty)}{T_e} = 1
	\end{align}
\end{subequations}
According to Mirels \cite{mirels1956boundary}, $T$ can be analytically expressed as the linear superposition of the solution without heat transfer plus the effect of heat transfer, i.e., 
\begin{subequations}
	\begin{align}
	\dfrac{T}{T_e} &= 1 + \left(\dfrac{u_w}{u_e}-1\right)^2\dfrac{{u^2_e} r\left(\eta\right)}{2T_e c_{p}}+ \left(\dfrac{T_w}{T_e}-\dfrac{T_r}{T_e}\right)s\left(\eta\right) \label{r} \\
	\dfrac{T_r}{T_e} &= 1 + \left(\dfrac{u_w}{u_e}-1\right)^2 \dfrac{u^2_e r\left(0\right)}{2T_e c_{p,w}} \label{s}
	\end{align}
\end{subequations}
where the dependent variables $r(\eta)$ and $s(\eta)$ are defined as follows
\begin{subequations}
	\begin{align}
	r'' + Pr_w f r' &= -\dfrac{2Pr_w}{\left(\dfrac{u_w}{u_e}-1\right)^2 }\left(f''\right)^2, \qquad 	r\left(\infty\right) = r'\left(0\right) = 0  \\
	s'' + Pr_w f s' &= 0, \qquad  \qquad  	s\left(0\right) =1 \qquad s\left(\infty\right) = 0
	\end{align} \label{ode-r-s}
\end{subequations}
The relation of the boundary layer displacement thickness $\delta^*$ can be then obtained as \cite{mirels1956boundary}
\begin{subequations}
	\begin{align}
	\delta^*(x') =  K_M\sqrt{\dfrac{\mu_e x'}{\rho_eu_e}} 
	\end{align}with the factor $K_M$ given by
	\begin{align}
	K_M = \sqrt{2\dfrac{\nu_w}{\nu_e}} \dfrac{\rho_w}{\rho_e}\abs{\lim_{\eta\to \infty}\left(\eta - f\right) + \left(\dfrac{u_w}{u_e}-1\right)^2\dfrac{{u^2_e}}{2T_e c_{p,w}}\intl{0}{\infty}rd\eta+ \left(\dfrac{T_w}{T_e}-\dfrac{T_r}{T_e}\right)\intl{0}{\infty}sd\eta} \label{KM}
	\end{align} 
\end{subequations}

\begin{figure}[]
	\centering
	{\includegraphics[width=0.5\textwidth]{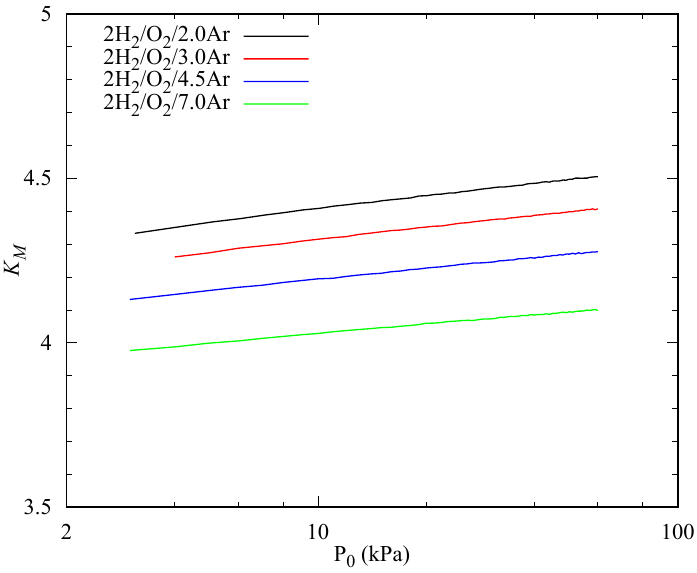}}
	\caption{Mirels' constant $K_M$ as a function of initial pressures for different mixtures.} \label{K-Laminar}  
\end{figure}

By solving the equations of Eq.\ (\ref{ode-f}) and Eq.\ (\ref{ode-r-s}) under the framework of Cantera and SDToolbox with the detailed San Diego reaction mechanism, we can obtain the Mirels' factor $K_M$ from Eq.\ (\ref{KM}) for different mixtures at varied initial pressures, which has been shown in Fig.\ \ref{K-Laminar}. It can be observed that the constant $K_M$ exhibits minor variations regarding the initial pressures and the mixture compositions. Note that the present realistic calculations using Cantera adopted the real thermodynamic data, while Mirels' model assumed the viscosity to scale linearly with temperature. Differences of the assumptions can introduce the variability of the evaluated $K_M$ by approximately 20\% to 30\%, when evaluating boundary layers behind detonations. In the present work, we adopt $K_M = 4.5$ for evaluating the laminar boundary layer displacement thickness \cite{mach2011mach, lau2019viscous}. 

\bibliographystyle{elsarticle-num}
\bibliography{mybibfile} 
\end{document}